\documentclass{aa}  
\usepackage{txfonts,epsfig,graphicx,url,twoopt}
\usepackage[breaklinks=true]{hyperref} 
\hypersetup{colorlinks=true,citecolor=blue,linkcolor=blue,urlcolor=red}
\graphicspath{{./figures/}}

\begin{document}

   \title{Close-in planetesimal formation by pile-up of drifting pebbles}

   \author{J. Dr{\k a}{\. z}kowska\inst{1}
          \and
          Y. Alibert\inst{2}
          \and
          B. Moore\inst{1}
          }

   \institute{Institute for Computational Science, University of Zurich,
              Winterthurerstrasse 190, 8057 Zurich, Switzerland\\
              \email{joannad@physik.uzh.ch}
           \and
              Physikalisches Institut \& Center for Space and Habitability, University of Bern, 
              Sidlerstrasse 5, 3012 Bern, Switzerland\\
             }

   \date{Received 23 May 2016 \slash~Accepted 17 July 2016}

 
  \abstract
   {The consistency of planet formation models suffers from the disconnection between the regime of small and large bodies. This is primarily caused by so-called growth barriers: the direct growth of larger bodies is halted at centimetre-sized objects and particular conditions are required for the formation of larger, gravitationally bound planetesimals.}
   {We aim to connect models of dust evolution and planetesimal formation to identify regions of protoplanetary discs that are favourable for the formation of kilometre-sized bodies and the first planetary embryos.}
   {We combine semi-analytical models of viscous protoplanetary disc evolution, dust growth and drift including backreaction of the dust particles on the gas, and planetesimal formation via the streaming instability into one numerical code. We investigate how planetesimal formation is affected by the mass of the protoplanetary disc, its initial dust content, and the stickiness of dust aggregates.}
   {We find that the dust growth and drift leads to a global redistribution of solids. The pile-up of pebbles in the inner disc provides local conditions where the streaming instability is effective. Planetesimals form in an annulus with its inner edge lying between 0.3~AU and 1~AU and its width ranging from 0.3~AU to 3~AU. The resulting surface density of planetesimals follows a radial profile that is much steeper than the initial disc profile. These results support formation of terrestrial planets in the solar system from a narrow annulus of planetesimals, which reproduces their peculiar mass ratios.}
   {}


   \maketitle

\section{Introduction}

Planets seem to be omnipresent in our Galaxy with most stars orbited by one or more of them \citep{2012Natur.481..167C}. However, we are a long way from a complete understanding of how these planets form. Despite recent progress in planet formation research, there is no model that can explain the growth of planets beginning with micron-sized dust grains. Most significantly, there is a disconnection between the models of the early stages of planet formation that are dealing with dust growth, and the late stages which follow the final accretion of planetary systems starting with planetesimals and embryos. The late stage models usually assume that planetesimals form rapidly with a smooth radial distribution \citep{2013apf..book.....A}. On the other hand, the early stage models typically end without producing any aggregates larger than cm-sized \citep[see e.g.][]{2014prpl.conf..339T}. This is because of so-called growth barriers that result from the collisional physics of dust aggregates and loss of solids because of their radial drift. The dust aggregates tend to bounce and fragment at the impact speeds predicted for the protoplanetary disc environment \citep{2010A&A...513A..57Z,2011A&A...525A..11B}. On the other hand, the radial drift timescale is shorter than the growth timescale for aggregates approaching centimetre sizes and the initial aggregates are removed from the disc before any larger bodies can form \citep{2008A&A...480..859B,2009A&A...503L...5B}. We give a concise overview of the typical predictions for dust evolution in the following section.

This paper is organised as follows. We give a brief introduction to the problem of growth barriers in Sect.~\ref{sub:barriers} and highlight the streaming instability as a possible solution in Sect.~\ref{sub:si}. In Sect.~\ref{sub:methods}, we explain the numerical approach that we employ to investigate dust evolution and planetesimal formation in a viscously evolving protoplanetary disc. We present results of our models in Sect.~\ref{sub:results}. Finally, we discuss limitations of our approach in Sect.~\ref{sub:discussion} and conclude the work in Sect.~\ref{sub:last}. For the readers convenience, the symbols used throughout this paper are summarised in Table~\ref{table:symbols}.

\subsection{Growth barriers}\label{sub:barriers}

\begin{figure*}
   \centering
   \includegraphics[width=0.8\hsize]{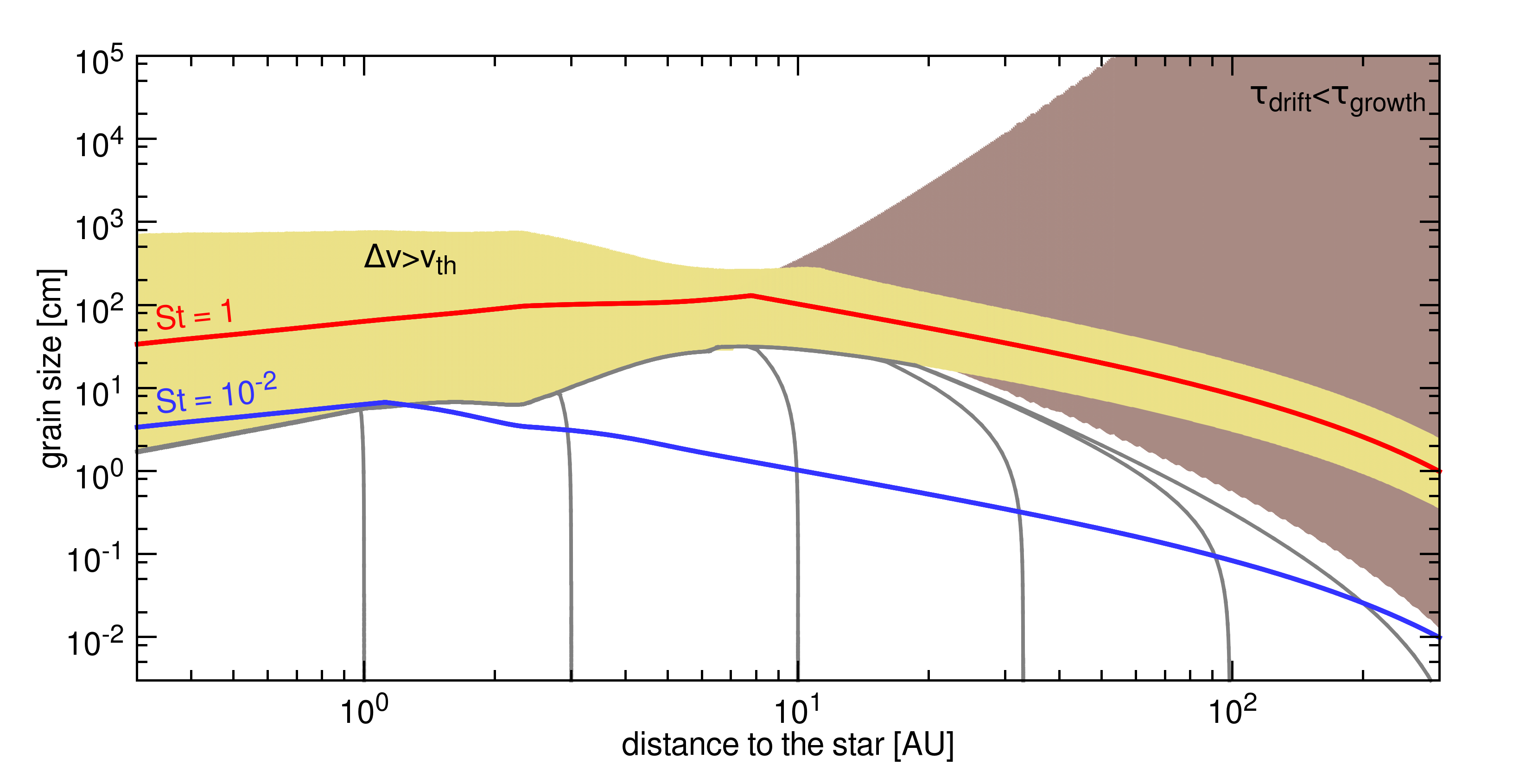}
      \caption{Overview of the growth barriers in our fiducial disc model. The red line corresponds to a Stokes number of unity. The yellow region marks the fragmentation barrier where the impact speeds would be higher then the threshold value of 10~m~s$^{-1}$. The grey region shows the size scale for which the drift timescale is shorter than the growth timescale, meaning that the aggregates would be removed from a given location faster than the growth could replenish them: this is the drift barrier. The light grey vertical lines are tracks of test particles evolving by growth and drift in a steady-state disc. Finally, the blue nearly horizontal line shows the size scale corresponding to the Stokes number of 10$^{-2}$, which is a minimum size required for planetesimal formation via the streaming instability.}
      \label{fig:barriersmap}
\end{figure*}

Dust evolution is driven by its interaction with the sub-Keplerian gas disc. The interaction between a dust aggregate and the surrounding gas can be conveniently parametrised with the so-called Stokes number
\begin{equation}\label{stokes1}
{\rm{St}} = t_{\rm{s}}\Omega_{\rm{K}},
\end{equation}
where $t_{\rm{s}}$ is the timescale over which the aggregate adjusts its velocity to the velocity of surrounding gas, and $\Omega_{\rm{K}}$ is Keplerian frequency. In this way, the Stokes number is the ratio between how quickly the aggregate reacts to the gas and the orbital period. Particles with $\rm{St}\ll1$ are tightly coupled to gas and $\rm{St}\gg1$ indicates decoupled solids. For our needs, it is convenient to rewrite Eq.~(\ref{stokes1}) as
\begin{equation}\label{stokes}
{\rm{St}} = \frac{\pi}{2}\frac{a\rho_{\bullet}}{\Sigma_{\rm{g}}},
\end{equation}
where $a$ is the radius and $\rho_{\bullet}$ is the internal density of the dust aggregate, and $\Sigma_{\rm{g}}$ is the surface density of the gas. Equation~(\ref{stokes}) applies to small and compact solids, where the radius $a$ is smaller than the mean-free path in gas. This is the so-called Epstein drag regime, where particles are well-coupled to the gas, and this regime is valid for all models presented in this paper. 

Figure~\ref{fig:barriersmap} gives an overview of global dust evolution in a protoplanetary disc with total mass of $0.1$~M$_{\odot}$, in terms of distance to the central star and dust aggregate size.
The size of aggregates corresponding to the Stokes number of unity, which is when they are affected by the interaction with the gas the most, is marked with the red nearly horizontal line. The blue line below shows a minimum size that is required for an efficient streaming instability (corresponding to ${\rm{St}}=10^{-2}$, see Sect.~\ref{sub:si}).
The outer part of the protoplanetary disc is dominated by the radial drift barrier (grey triangle-shaped region), where the growth timescale 
\begin{equation}\label{tgrowth}
\tau_{\rm growth} = a~\left(\frac{{\rm d}a}{{\rm d}t}\right)^{-1} = \frac{a\rho_{\bullet}}{\rho_{\rm{d}} \Delta v}
\end{equation}
is longer than the radial drift timescale
\begin{equation}\label{tdrift}
\tau_{\rm drift} = r \left(\frac{{\rm d}r}{{\rm d}t}\right)^{-1} = \frac{r}{|v_{\rm r,d}|},
\end{equation}
where $\rho_{\rm{d}}$ is the density of dust, $\Delta v$ is the impact velocity for collisions between dust aggregates, $r$ is the radial distance to the star, and $v_{\rm r,d}$ is the radial drift velocity. 
We assume that the growth preferentially happens between equal-sized grains that have settled to the midplane when deriving Eq.~(\ref{tgrowth}). This means that the dust density may be written as 
\begin{equation}\label{rhod}
\rho_{\rm{d}} = \frac{\Sigma_{\rm{d}}}{\sqrt{2\pi}H_{\rm{d}}},
\end{equation}
where $\Sigma_{\rm{d}}$ is the dust surface density and $H_{\rm{d}}$ is the scale height of the dust, which depends on the turbulence strength $\alpha_{\rm{t}}$ and the Stokes number of grains $\rm{St}$. In the case of a single-sized population
\begin{equation}\label{Hd}
H_{\rm{d}} = H_{\rm{g}} \sqrt{\frac{\alpha_{\rm{t}}}{\alpha_{\rm{t}}+{\rm{St}}}},
\end{equation}
where $H_{\rm{g}} = c_{\rm{s}} \Omega_{\rm{K}}^{-1}$ is the scale height of the gas. 

We note that the growth timescale $\tau_{\rm{growth}}$ given by Eq.~(\ref{tgrowth}) simplifies to 
\begin{equation}\label{tgrowth2}
\tau_{\rm{growth}} \approx \frac{1}{Z\cdot\Omega_{\rm K}}
\end{equation}
under the assumptions that the collisions are driven by turbulence, in which case the impact speed $\Delta v \approx \sqrt{3\alpha_{\rm t}{\rm St}}\cdot c_{\rm s}$ \citep{2007A&A...466..413O}, the dust grains are in the Epstein regime (see Eq.~\ref{stokes}), and the dust density in the midplane is given by Eq.~(\ref{rhod}). We emphasise that, because $\Delta v \propto {\alpha_{\rm t}}^{1 / 2}$ and $\rho_{\rm d} \propto {\alpha_{\rm t}}^{-1 / 2}$, the dependence of the dust growth timescale on the turbulence strength parameter $\alpha_{\rm t}$ cancels itself out.

This initial growth stage is significant particularly at large orbital distances, where the growth is much slower than in the inner parts of the protoplanetary disc. This results in a delayed delivery of pebbles from the outer disc to its inner regions. The same effect was also described by \citet{2014A&A...572A.107L}, who called it a "pebble formation front".

The radial drift limits the maximum size of grains that are available in the outer parts of the disc. The pebbles that grow far away from the star are then shifted to the inner disc, where they undergo fragmentation during high-speed collisions. We assume that the fragmentation happens when the collision speed exceeds a threshold value, which we set to $v_{\rm{th}}=10$~m~s$^{-1}$ for Fig.~\ref{fig:barriersmap}. 

The evolution of dust in the outer disc is determined by the interplay between growth and drift. Because the timescale of drift is shorter than what is needed to grow to centimetre sizes at a few tens of AU from the central star, this outer region is gradually depleted on a timescale of a few Myrs. 
However, as may be seen in Fig.~\ref{fig:barriersmap}, the radial drift barrier does not stretch all the way to the inner edge of the disc. In the inner disc, the collisional timescale is shorter than the drift timescale. If growth happens even at high impact speeds, for example in the case of very porous dust aggregates, planetesimal formation via direct growth may be possible \citep{2012ApJ...752..106O, 2013A&A...557L...4K}. Even for compact grains, interplay between the radial drift and dust growth can lead to a pile-up of solids in the inner regions of the disc \citep{2012A&A...539A.148B, 2012A&A...537A..61L, 2014A&A...565A.129P}. This pile-up is required by the planetesimal formation models which include the streaming instability \citep{2009ApJ...704L..75J,2010ApJ...722.1437B}. At the same time, the streaming instability requires the presence of relatively large pebbles, with sizes corresponding to the Stokes number of $\rm{St}>10^{-2}$ \citep{2014A&A...572A..78D}. This is not necessarily the case because of the bouncing and fragmentation barriers. However, as can be seen in Fig.~\ref{fig:barriersmap}, there is a region around 1-10~AU where the maximum size of grains that can be reached because of fragmentation is above the $\rm{St}=10^{-2}$ line and the radial drift barrier is not efficient. In this paper, we check whether the redistribution of solids driven by the radial drift and growth in a realistic viscous disc can lead to planetesimal formation via the streaming instability.

\subsection{Streaming instability}\label{sub:si}

Owing to the growth barriers described in the previous section, direct growth from micron to km-sizes seems unlikely. Streaming instability that is able to produce planetesimals directly out of cm-sized pebbles is a good solution to the planetesimal formation issue \citep{2007Natur.448.1022J}. 

However, planetesimal formation via the streaming instability requires enhancement of the dust-to-gas ratio by a factor of a~few over the standard solar value of 10$^{-2}$ \citep{2009ApJ...704L..75J, 2010ApJ...722.1437B}. There are different scenarios that modify disc structure and introduce pressure bumps that make it possible to obtain such enhancements: the zonal flows \citep{2011A&A...529A..62J, 2013ApJ...763..117D}, vortices \citep{2015ApJ...804...35R, 2016arXiv160105945S}, dead zone edges \citep{2009A&A...497..869L}, and planet-disc interactions \citep{2012MNRAS.423.1450A}. In the models presented in this paper, we check whether the streaming instability can work thanks to a dust pile-up induced in the inner disc by the combination of growth and drift. This scenario does not require any ad-hoc assumptions about the disc structure or pre-existing planets. 

Until now the streaming instability was only modelled in local or quasi-global simulations. We use a global 1D semi-analytical protoplanetary disc model together with a prescription for streaming instability extracted from the local hydrodynamic simulations, similar to \citet{2014A&A...572A..78D} for a local case. As a consequence, we are able to identify regions of the disc in which planetesimal formation happens. We perform an extended parameter study to investigate how the planetesimal formation is affected by the disc mass, metallicity, and stickiness of dust aggregates. 
The modelling methods that we use are described in Sect.~\ref{sub:methods}. 

\section{Methods}\label{sub:methods}

In our models, we take into account the viscous evolution of protoplanetary discs, dust growth and drift, as well as planetesimal formation via the streaming instability. We model how the surface density of gas, dust, and planetesimals evolves over timescales of up to 10 Myrs. 
Our approach combines semi-analytical models of viscous protoplanetary disc evolution that follows the work of \citet{2005A&A...434..343A, 2013A&A...558A.109A}, dust evolution model based on the approach proposed by \citet{2012A&A...539A.148B}, and planetesimal formation via the streaming instability similar to \citet{2014A&A...572A..78D}. The following subsections describe each of these components. 

\subsection{Disc model}\label{sub:gasdisc}

\begin{figure}
   \centering
   \includegraphics[width=0.95\hsize]{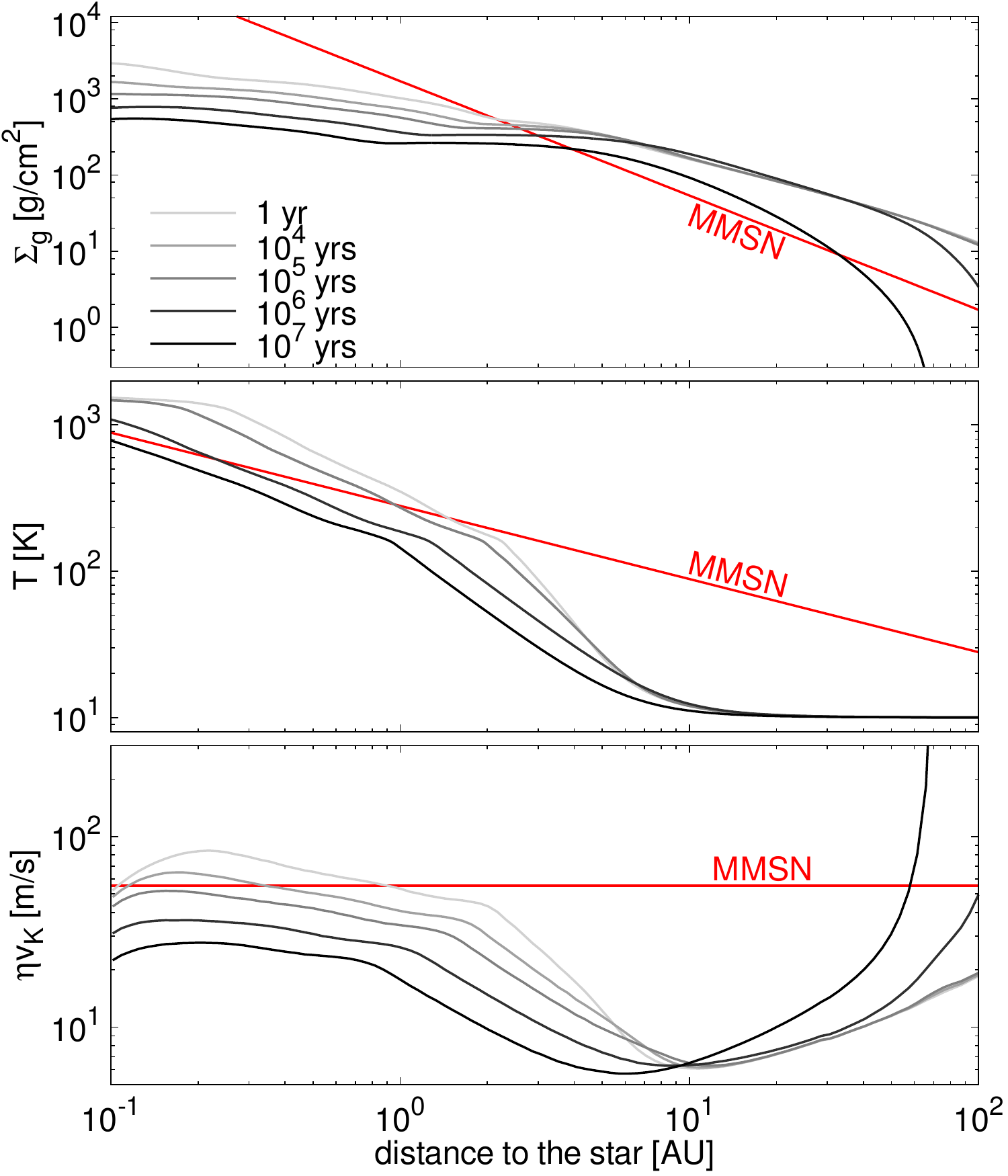}
      \caption{Time evolution of the gas surface density $\Sigma_{\rm{g}}$, gas temperature, and the headwind speed $\eta v_{\rm{K}}$ in our fiducial disc model. The disc mass is initially $0.1$~M$_{\odot}$ and it decreases to $0.02$~M$_{\odot}$ after 10~Myrs. The disc is shallower then the Minimum Mass Solar Nebula (MMSN, red line) and thus most of its mass resides in its outer parts. The headwind, which determines the maximum radial drift velocity, is not constant throughout the disc, as opposed to the MMSN model, and it gradually decreases during the evolution, facilitating the retention of solids.}
      \label{fig:gasevo}
\end{figure}

Our gas disc model is the same as that used in \citet{2005A&A...434..343A, 2013A&A...558A.109A} and is computed using the method originally derived by \citet{1999ApJ...521..823P}. Here, we give a general description of the model and refer the reader to the above-mentioned papers for further details.

The evolution of the disc is computed using a 1+1D model in a two-step process. In the first step, we compute the vertical structure of the disc for each distance to the star by solving the equation of hydrostatic equilibrium, energy conservation (taking into account viscous dissipation), and radiative diffusion. For this calculation, the opacity is assumed to be that of \citet{1994ApJ...427..987B}, which is representative of the opacity in the interstellar medium. In the calculations present in the paper, we did not include the effect of stellar irradiation. The viscosity is assumed to follow the well-known Shakura-Sunyaev parametrisation \citep{1973A&A....24..337S}, with an $\alpha_{\rm t}$ parameter equal to $10^{-3}$. For higher $\alpha_{\rm{t}}$ values, it is very hard to obtain a dense midplane layer of dust that is necessary to trigger the streaming instability (see Sect.~\ref{sub:pf}). The solution of the vertical structure equation gives the distribution of temperature, pressure and radiative flux along the vertical axis for each distance to the star and as a function of the gas surface density, as well as the mass-weighted mean viscosity. This latter quantity is finally used, in the second step, to solve the radial diffusion equation that provides the evolution of the gas surface density as a function of time. This diffusion equation was modified to include far-ultraviolet (FUV) photoevaporation following the prescription of \citet{2004MNRAS.347..613V} with the mass-loss rate of $\dot{M}_{\rm{wind}}=10^{-6}$~M$_{\odot}$~yr$^{-1}$. With this prescription, photoevaporation is significant particularly at large orbital distances where the gas is less bound, and the viscous spreading of the gas disc is suppressed.

Figure \ref{fig:gasevo} highlights the evolution of gas in our fiducial model. The initial mass of this disc is $0.1$~M$_{\odot}$, which decreases to $0.02$~M$_{\odot}$ after 10~Myrs. 
The disc cools with time, which modifies the impact speeds of dust aggregates, as well as the pressure structure of the disc and thus the radial drift. 
The headwind $\eta v_{\rm{K}}$, which determines the maximum radial drift speed, is set by the gas pressure $P_{\rm{g}}$ gradient
\begin{equation}\label{etavk}
\eta v_{\rm{K}} = \frac{1}{2\rho_{\rm{g}}\Omega_{\rm{K}}} \frac{dP_{\rm{g}}}{dr},
\end{equation}
where $\rho_{\rm{g}}$ is gas density and $\Omega_{\rm{K}}$ is Keplerian frequency. The pressure gradient is negative and the negative values of velocity translate into inward drift. The maximum drift speed decreases during the evolution. This facilitates the retention of solids and growth of the dust pile-up in the inner part of the disc (see Sect.~\ref{sub:results}). 

\subsection{Dust evolution}\label{sub:dustevo}

Models that include both dust advection and growth, as pioneered by \citet{1997Icar..127..290W}, are computationally demanding and their results may be difficult to interpret owing to their complexity. We focus on simplified models, only including the necessary physics in a semi-analytical manner since we want to gain an overview of the dust evolution process and understand its connection to the planetesimal formation stage.  

Instead of directly modelling collisions between dust grains, we prescribe their growth and fragmentation using a prescription similar to the one proposed by \citet{2012A&A...539A.148B}. We describe the main aspects of this approach here and refer the interested readers to the full paper.

The approach is based on solving the advection-diffusion equation for dust surface density $\Sigma_{\rm d}$
\begin{equation}\label{advdiff}
\frac{\partial \Sigma_{\rm d}}{\partial t} + \frac{1}{r} \frac{\partial}{\partial r}\left[r\left(\Sigma_{\rm d}\bar{v}-D_{\rm g}\Sigma_{\rm g}\frac{\partial}{\partial r}\left(\frac{\Sigma_{\rm d}}{\Sigma_{\rm g}}\right)\right)\right] = 0,
\end{equation}
where $r$ is distance to the central star, $\Sigma_{\rm g}$ is the surface density of gas, and $D_{\rm g}$ is gas diffusivity.
The dust advection speed $\bar{v}$ is a mass-weighted average (see Eq.~\ref{mwvel}). The advection velocity corresponding to each Stokes number is described by Eq.~(\ref{vmnsh}). This consists of two components: one corresponding to the difference between gas and Keplerian rotation and one related to the radial velocity of gas. The radial drift speed depends on the maximum size of dust aggregates and on the size distribution. The size distribution is regulated by the processes that limits growth at given location: fragmentation or radial drift (see~Sect.~\ref{sub:barriers}). 

If the maximum size of dust aggregates is limited by the radial drift, then we assume that the size distribution is very narrow and focussed around
\begin{equation}\label{adrift}
a_{\rm{drift}} = {\rm{f_d}} \frac{2}{\pi} \frac{\Sigma_{\rm{d}}}{\rho_{\bullet}}\frac{v_{\rm{K}}^2}{c_{\rm{s}}^2}\left|\frac{d\ln{P_{\rm{g}}}}{d\ln{r}}\right|^{-1},
\end{equation}
where $v_{\rm{K}}$ is the Keplerian velocity, $c_{\rm{s}}$ is sound speed, and $\rm{f_d}=0.55$ is a model parameter calibrated by \citet{2012A&A...539A.148B}.
In the inner part of the disc, the size distribution is regulated by the equilibrium between coagulation and fragmentation \citep{2011A&A...525A..11B} and it takes the form of a power law described by Eq.~(\ref{sizedistr}). In this case, we assume that the maximum size is
\begin{equation}\label{afrag}
a_{\rm{frag}} = {\rm{f_f}} \frac{2}{3\pi} \frac{\Sigma_{\rm{g}}}{\rho_{\bullet}\alpha_{\rm t}}\frac{v_{\rm{th}}^2}{c_{\rm{s}}^2},
\end{equation}
where $v_{\rm{th}}$ is the fragmentation threshold velocity and the calibration factor $\rm{f_f} = 0.37$. In all the models presented in this paper, fragmentation is driven by the turbulent velocities rather than differential drift. If the $\alpha_{\rm{t}}$ parameter would be very low, i.e. in a disc with dead zone, then Eq.~(\ref{afrag}) would have to be modified to take the differential drift into account.
The sizes indicated by Eqs.~(\ref{adrift}) and (\ref{afrag}) correspond to the lower boundaries of the growth-barrier regions depicted in Fig.~\ref{fig:barriersmap}.

We assume that at the beginning of the evolution all the dust grains have a size of $a_0$, independent of orbital distance. For this paper, we adopt $a_0=1$~$\mu$m. The dust grows until it reaches the size limited by fragmentation or radial drift. The growth is prescribed such that
\begin{equation}\label{a1}
a(t) = \min\left(a_0\cdot\exp\left(t\slash\tau_{\rm{growth}}\right),a_{\rm frag},a_{\rm drift}\right),
\end{equation}
where the first condition accounts for the initial growth phase and the growth timescale $\tau_{\rm{growth}}$ is described by Eq.~(\ref{tgrowth2}). 

As it was demonstrated by \citet{2012A&A...539A.148B}, this type of simplified approach is able to fit the results of full dust evolution simulations surprisingly well, while greatly reducing the computational cost. 

One major difference between the code of \citet{2012A&A...539A.148B} and ours is that we take into account the so-called collective effects, or the effects of backreaction of dust on gas, while calculating the radial drift velocity $\bar{v}$. \citet{2010ApJ...722.1437B} noticed that the radial drift is modified when clumping occurs in their streaming instability models. The rate of radial drift decreases in a way that depends both on the local dust-to-gas ratio and on the size distribution of grains. This is because the different dust species are indirectly coupled owing to their interaction with gas. We implement these effects using equations derived by \citet{2005ApJ...625..414T}, which are also found in \citet{2012ApJ...752..106O}. Details of our implementation are presented in Appendix~\ref{sub:mNSH}.

This modification of radial drift significantly strengthens the effects of dust overdensities. If there is a location where the dust-to-gas ratio is enhanced, the drift slows down and a pile-up of material arises. We demonstrate the importance of the backreaction on the global dust retention in Fig.~\ref{fig:massevo} (the black dashed line versus the grey solid line).

In contrast to \citet{2012A&A...539A.148B}, who employed an implicit integration scheme, we perform explicit integration of Eq.~(\ref{advdiff}) using a total variation diminishing scheme and employing the input provided by the gas evolution module. We limit the time step to
\begin{equation}\label{dtdust}
dt = \min{\left(C\frac{dr}{|\bar{v}|},C\frac{dr^2}{D_{\rm g}}\right)},
\end{equation}
where $dr$ is width of radial grid cell, and $C<1$ to assure numerical accuracy. Explicit integration is necessary to add planetesimal formation into this model, as is described in the following section. 

\subsection{Planetesimal formation}\label{sub:pf}

As mentioned in Sect.~\ref{sub:si}, we account for planetesimal formation via the streaming instability using a semi-analytic model similar to \citet{2014A&A...572A..78D}. However, they focussed on local models and studying the connection between dust growth and planetesimal formation. The planetesimals were formed out of the reservoir of pebbles that has been replenished by growth and the loss of pebbles because of radial drift was neglected. In this paper, we extend this approach to a global disc model, including both dust growth (albeit in a simplified fashion as discussed in the previous section) and the radial drift. The radial drift has both positive and negative effects on planetesimal formation: it makes planetesimal formation possible by creating pile-ups of dust in the inner disc and then it delivers the pebbles from the outer disc, extending the planetesimal formation period. However, it also gradually removes solids from the disc, restricting the amount of planetesimals that can be formed. 

Another major difference between our approach and that of \citet{2014A&A...572A..78D} is that they focussed on dead zones, where turbulence was only present when triggered by the midplane instability caused by dust sedimentation. In this paper, we focus on a viscous disc model, where external turbulence is present and it is parametrised by $\alpha_{\rm t}=10^{-3}$. This causes a change to the dominant condition determining the possibility of planetesimal formation. \citet{2014A&A...572A..78D} defined two conditions that have to be simultaneously fulfilled to allow for planetesimal formation:
\begin{itemize}
\item{the vertically integrated dust-to-gas ratio of pebbles with sizes corresponding to $\mathit{St} > 10^{-2}$ exceeds a critical value $Z_{\rm crit}$,}
\item{the dust-to-gas ratio in the midplane exceeds unity.}
\end{itemize}
The value of $Z_{\rm crit}$ was calibrated on the hydrodynamical simulations results presented by \citet{2010ApJ...722.1437B,2010ApJ...722L.220B} who considered a laminar disc. Similar results were presented recently by \citet{2015A&A...579A..43C}, who also included dependence of the $Z_{\rm crit}$ on the Stokes number of particles. Here, with the external turbulence present, the vertically integrated dust-to-gas ratio $\Sigma_{\rm d}/\Sigma_{\rm g}$ has to be much higher than $Z_{\rm crit}$ used in previous work to form a dense midplane layer of solids. In other words, the second condition pointed out by \citet{2014A&A...572A..78D} is always more restrictive than the first one. This is why we do not use the $Z_{\rm crit}$, but rather focus on the midplane dust-to-gas ratio in this paper. Following Eq.~(\ref{Hd}), the relation between the vertically integrated and the midplane dust-to-gas ratio in the case of a single dust size is
\begin{equation}\label{dtg}
\frac{\rho_{\rm d}}{\rho_{\rm g}} = \frac{\Sigma_{\rm d}}{\Sigma_{\rm g}}\sqrt{\frac{\alpha_{\rm t}+\mathit{St}}{\alpha_{\rm t}}},
\end{equation}
where $\rho_{\rm d}$ and $\rho_{\rm g}$ are, respectively, the dust and gas midplane density.

The planetesimal formation algorithm we implement works as follows. In each timestep and in each grid cell, we check whether the condition for planetesimal formation is fulfilled, namely that the density of sufficiently large pebbles in the midplane is comparable to the gas density
\begin{equation}\label{condition}
\sum_{\rm St>10^{-2}}\frac{\rho_{\rm d}({\rm St})}{\rho_{\rm g}} > 1.
\end{equation}
If this is the case, we transfer part of the surface density of dust into planetesimals, such that
\begin{equation}\label{sigmaplts}
d\Sigma_{\rm{plts}} = \zeta\frac{\Sigma_{\rm d}({\rm St}>10^{-2})}{T_{\rm K}} dt,
\end{equation}
where $T_{\rm K}$ is orbital period and $\zeta$ is planetesimal formation efficiency, that is the ratio of sufficiently large pebbles that are turned to planetesimals within one orbital period. We test different values of $\zeta$ in Sect.~\ref{sub:zeta}.

\section{Results}\label{sub:results}
\subsection{Fiducial model}

\begin{figure}
   \centering
   \includegraphics[width=\hsize]{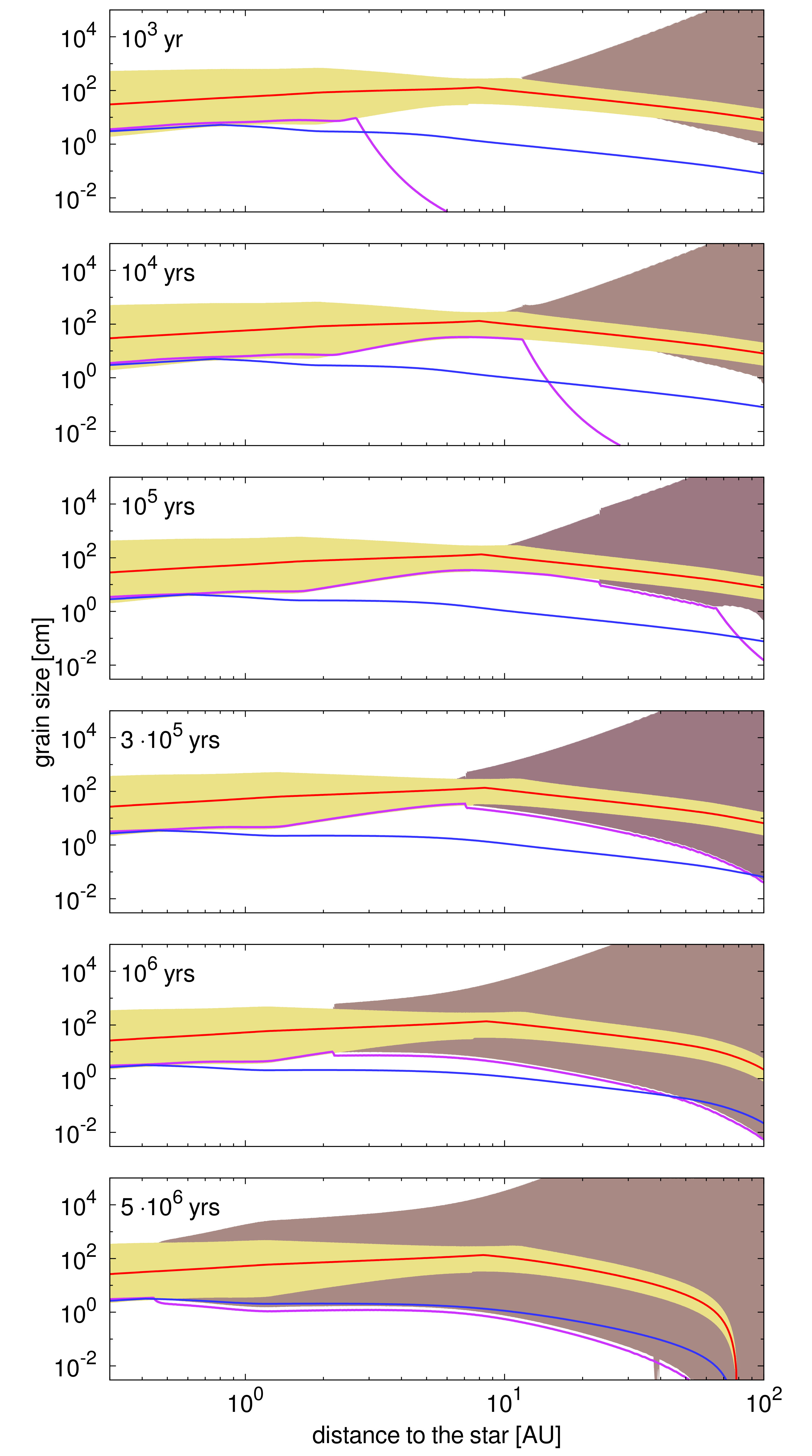}
      \caption{Time evolution of the growth barriers. The plotting style corresponds to Fig.~\ref{fig:barriersmap}: the yellow region corresponds to the fragmentation barrier while the grey region indicates the drift regime. The red and blue horizontal lines denote sizes corresponding to the Stokes number of unity and 10$^{-2}$, respectively. The purple line shows the maximum size of dust aggregates, which generally follows the growth barriers, except for the earliest stages of evolution, when it is restricted by the growth timescale. The radial drift barrier gains significance as the dust-to-gas ratio decreases during the evolution.}
      \label{fig:6panels}
\end{figure}

For our reference run, we choose a protoplanetary disc of mass M$_{\rm disc}=$~0.1~M$_\odot$, global dust-to-dust ratio of $Z=10^{-2}$, and turbulence strength parametrised by $\alpha_{\rm t}=10^{-3}$. We set the fragmentation threshold to $v_{\rm th}=10$~m~s$^{-1}$, independent of orbital distance, and the planetesimal formation efficiency parameter $\zeta=10^{-4}$. As our initial condition, we assume a constant $\Sigma_{\rm d} \slash \Sigma_{\rm g} = Z$ for the radial distance between 0.3~AU and 100~AU and $\Sigma_{\rm d}=0$ otherwise.

Figure~\ref{fig:6panels} shows the evolution of dust sizes restricted by growth barriers during our fiducial run. The combined action of dust growth and drift leads to the gradual loss of solids from the disc. Because of this, the importance of drift barriers is modified and the radial drift barrier strengthens with time (the grey triangle-shaped region moves towards the star). The fragmentation barriers slightly weakens because the gas temperature drops (see Fig.~\ref{fig:gasevo}). The representative size of dust follows the restrictions imposed by growth barriers, except for the beginning of the simulation, when it is determined by dust growth. 

\begin{figure}
   \centering
   \includegraphics[width=0.95\hsize]{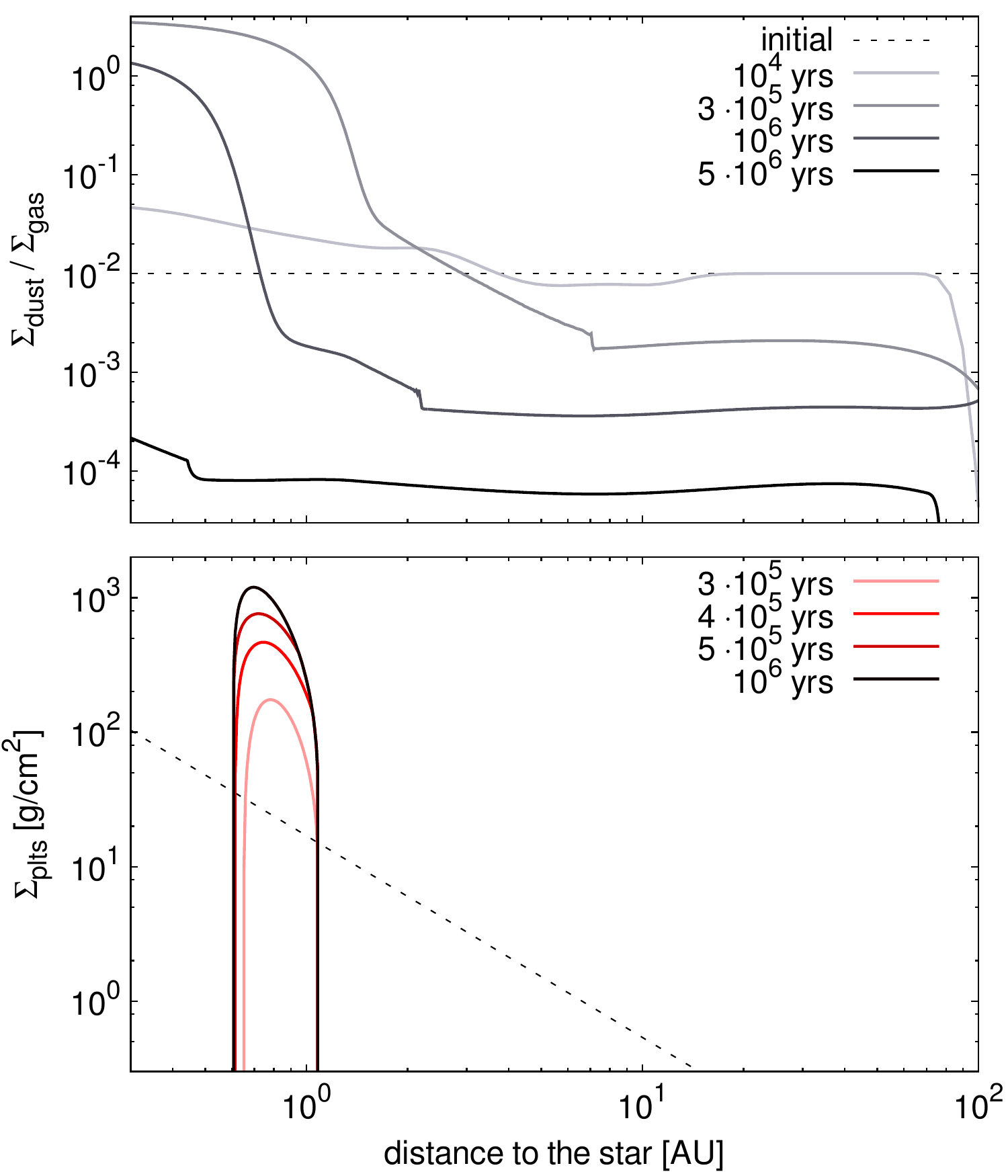}
      \caption{Time evolution of a)~the vertically integrated dust-to-gas ratio and b)~surface density of planetesimals. The outer parts of the disc are depleted of dust because they are drift dominated. At the same time, there is a temporary pile-up of dust in the fragmentation-dominated inner disc, which enables planetesimal formation via the streaming instability. The dashed line in panel b) corresponds to the surface density of solids in the MMSN model.}
      \label{fig:surfacedens}
\end{figure}

\begin{figure}
   \centering
   \includegraphics[width=\hsize]{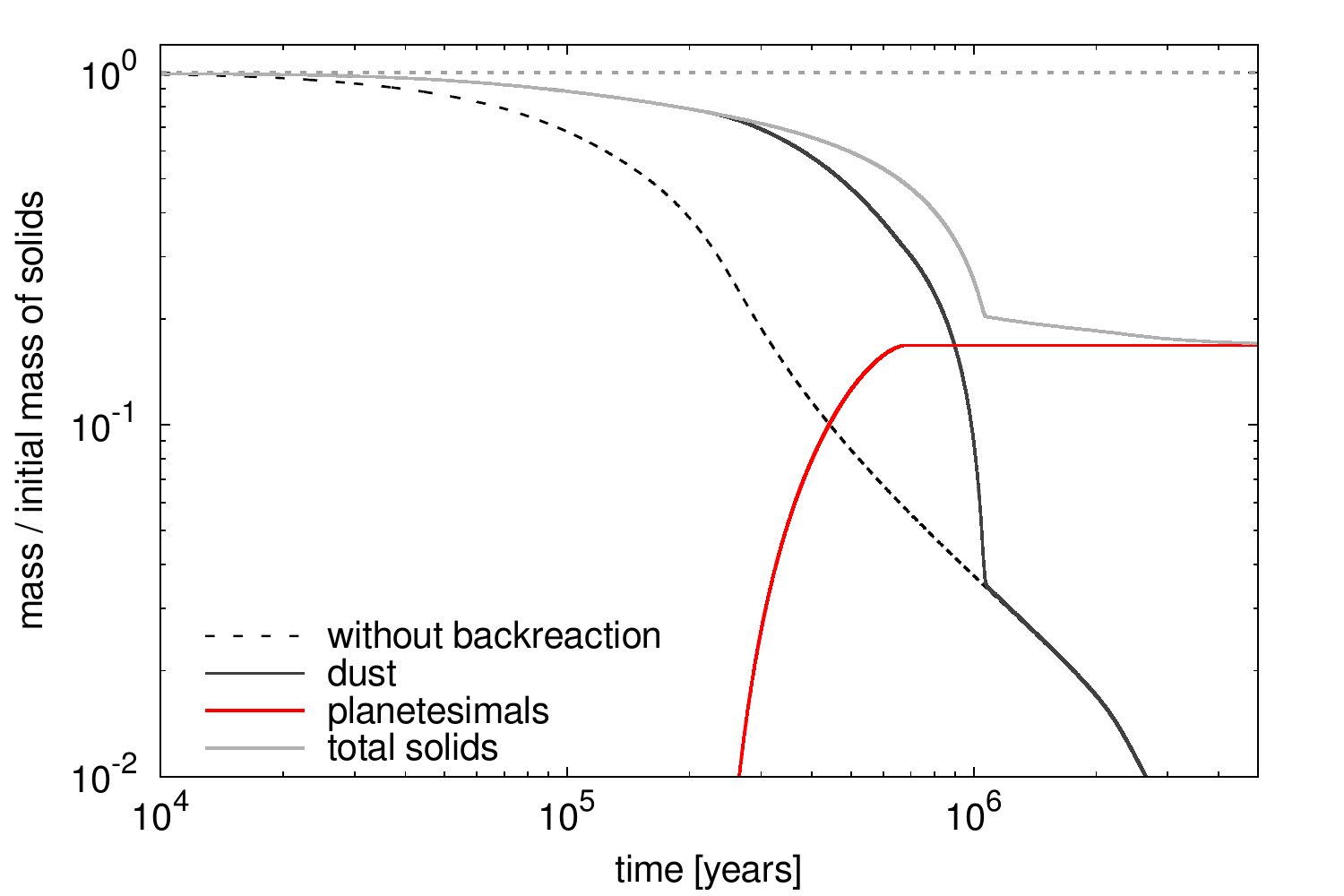}
      \caption{Retention of solids in our fiducial model and in a model without backreaction of solids on the gas flow. Initially all the solids are in the form of dust (dotted line at 1). Solids are lost primarily because of the radial drift. If the radial drift is independent of the dust-to-gas ratio ("without backreaction", black dashed line), $99\%$ of solids are lost within 2 Myrs and no planetesimals are formed. In the model presented in this paper that includes the backreaction, the radial drift is slowed down when dust piles up. This allows for planetesimal formation (red solid line) which keeps around $17\%$ of the initial mass of solids (light grey solid line) after the rest of the disc disperses (dark grey solid line).}
      \label{fig:massevo}
\end{figure}

In its outer, drift-dominated part, the dust density gradually decreases, but the interplay between dust growth and drift leads to the formation of a temporary pile-up in the inner disc (upper panel of Fig.~\ref{fig:surfacedens}). The formation of this pile-up is aided by the disc dispersal. As can be seen in the bottom panel of Fig.~\ref{fig:gasevo}, the maximum drift speed decreases with time in the inner part while increasing in the outer. This means that less and less solids flow out through the inner edge of the disc, which we set to 0.3~AU for dust, based on the evaporation temperature of silicates, while more and more dust arrives to the inner part from the outside.

As mentioned in Sect.~\ref{sub:dustevo}, the radial drift velocity in our models includes the so-called collective effects or, in other words, the effects of backreaction of dust on gas, that becomes important when dust-to-gas ratio increases. The drift slows down when the dust-to-gas ratio increases. Because of this, the initial pile-up becomes stronger, finally leading to conditions that are sufficient for triggering planetesimal formation via the streaming instability (see Sect.~\ref{sub:pf}) after $\sim 3 \cdot 10^{-5}$~yrs (lower panel of Fig.~\ref{fig:surfacedens}). 

As the initial disc profile is shallow, corresponding to roughly $\Sigma_{\rm g}\propto r^{-1}$, at the beginning of the simulation most of the mass of solids is located in the outer part of the disc. As the dust growth timescale is much longer at large orbital radii, these solids are then delivered to the inner disc later, which feeds the zone where planetesimal formation is possible and extends the planetesimal formation period until about 1~Myr of evolution. Afterwards, the inner disc becomes too depleted and planetesimal formation stalls.

The gradual loss of solids, as well as planetesimal formation, are depicted in Fig.~\ref{fig:massevo}. In our fiducial model, the total mass of solids (light grey solid line), split between dust (dark grey solid line) and planetesimals (red solid line), never actually decreases to zero and this is because about 17\% of the initial dust mass, which corresponds to 60 Earth masses, is turned into planetesimals. The planetesimals do not undergo radial drift. For comparison, we show the evolution of solids in the case when the the effects of backreaction on the radial drift speed are not included (black dashed line). No planetesimals are formed and all the solids are lost in this case.

The amount of planetesimals formed, as well as the location of the annulus in which they form, changes with the initial conditions. In the following sections, we describe the dependence of these results on some of the parameters: disc mass $M_{\rm disc}$, initial dust-to-gas ratio $Z$, fragmentation threshold velocity $v_{\rm th}$, and planetesimal formation efficiency $\zeta$. All these results are summarised in Fig.~\ref{fig:4panels} and Table~\ref{table:allruns}.

\subsection{Dependence on disc mass}

Changing the mass of the disc alters the growth barriers (Fig.~\ref{fig:barriersmap}) and thus the dust evolution. In a lower mass disc, the relative importance of the radial drift regime increases, moving the inner pile-up region even closer to the star (panel~a in Fig.~\ref{fig:4panels}). Fortunately, at the same time the lower mass discs are colder and, because to this, the maximum size of grains in the fragmentation-limited case increases (see Eq.~\ref{afrag}). This makes it possible to grow aggregates to the pebble size and still form planetesimals, although much closer to the star. Surprisingly, we find that the mass of planetesimal annulus relative to the initial mass of solids in the disc does increase with decreasing disc mass (6th column of Table~\ref{table:allruns}). This is because the lower mass discs disperse faster, facilitating the inner pile-up. However, the absolute mass of planetesimals decreases and is as low as six Earth masses for the disc with initial mass of 0.005~$M_{\odot}$. For the high mass discs, planetesimal formation becomes harder as the fragmentation barrier does not allow the formation of sufficiently large  pebbles. No planetesimals are formed in discs more massive than $0.2$~M$_\odot$.

\subsection{Dependence on metallicity}

Higher initial dust-to-gas ratios makes planetesimal formation much easier and more efficient. This is not only because there are more solids available, but also because the radial drift regime is pushed outwards as the growth timescale is shorter for higher dust densities (see Eq.~\ref{tgrowth}). We find that the solar metallicity of $Z=10^{-2}$ is the lowest that allows planetesimal formation (while keeping all the other parameters at their default values). The total mass of planetesimals that form increases linearly with the initial mass of solids, reaching $1400$ Earth masses for a disc that initially contains 5\% solids. At the same time, the outer edge of the planetesimal annulus moves outward, so the planetesimal formation region is broader, reaching 3~AU for $Z=0.05$. These results should naturally lead to the well-known metallicity-giant planet occurrence correlation \citep{2007ARA&A..45..397U}.

\subsection{Dependence on fragmentation threshold}\label{sub:vth}

We find that the aggregates should be able to grow at impact velocities of up to 8~m~s$^{-1}$ to allow planetesimal formation. This is because sufficiently large pebbles need to be produced and the maximum size of grains depends strongly on the fragmentation threshold (see Eq.~\ref{afrag}). On the other hand, we observe that planetesimal formation does not occur if the aggregates are allowed to grow at velocities above $v_{\rm th}=15$~m~s$^{-1}$. This is because the aggregates with Stokes numbers close to unity, which are formed with such a high fragmentation threshold, drift very quickly and are not able to form a significant pile-up. 

The properties of the planetesimal annulus change with $v_{\rm th}$. For higher velocities, the annulus moves inwards because of the faster radial drift. The mass of planetesimals that form slightly increases with higher $v_{\rm th}$ values, reaching 69 Earth masses for $v_{\rm th}=11$~m~s$^{-1}$, and decreases to 12 Earth masses for $v_{\rm th}=15$~m~s$^{-1}$.

\begin{table}
\caption{Summary of parameters used and results obtained in different models}
\centering                         
\begin{tabular}{l l l l l l l}   
\hline\hline                
ID & $M_{\rm{disc}}${\slash}$M_{\star}$ & $Z$ & $v_{\rm th}$\tablefootmark{a} & $\zeta$ & $M_{\rm plts}${\slash}$M_{\rm tot}$ & $M_{\rm plts}$\tablefootmark{b}\\   
\hline
01 & 0.1 & 0.01 & 10 & $10^{-4}$ & 0.17 & 60 \\
\hline
02 & 0.005 & 0.01 & 10 & $10^{-4}$ & 0.34 & 6 \\
03 & 0.01 & 0.01 & 10 & $10^{-4}$ & 0.32 & 11.2 \\
04 & 0.02 & 0.01 & 10 & $10^{-4}$ & 0.33 & 22.8 \\
05 & 0.03 & 0.01 & 10 & $10^{-4}$ & 0.32 & 32.3 \\
06 & 0.05 & 0.01 & 10 & $10^{-4}$ & 0.27 & 45.1 \\
07 & 0.08 & 0.01 & 10 & $10^{-4}$ & 0.21 & 56.3 \\
08 & 0.15 & 0.01 & 10 & $10^{-4}$ & 0.11 & 52.7 \\
09 & 0.2 & 0.01 & 10 & $10^{-4}$ & 0.01 & 9.2 \\
10 & 0.3 & 0.01 & 10 & $10^{-4}$ & 0 & 0 \\
\hline
11 & 0.1 & 0.005 & 10 & $10^{-4}$ & 0 & 0 \\
12 & 0.1 & 0.014 & 10 & $10^{-4}$ & 0.37 & 182.4 \\
13 & 0.1 & 0.02 & 10 & $10^{-4}$ & 0.53 & 376.5 \\
14 & 0.1 & 0.03 & 10 & $10^{-4}$ & 0.66 & 710.6 \\
15 & 0.1 & 0.04 & 10 & $10^{-4}$ & 0.74 & 1050.1 \\
16 & 0.1 & 0.05 & 10 & $10^{-4}$ & 0.78 & 1392.4 \\
\hline
17 & 0.1 & 0.01 & 8 & $10^{-4}$ & 0 & 0 \\
18 & 0.1 & 0.01 & 9 & $10^{-4}$ & 0.07 & 24.4 \\
19 & 0.1 & 0.01 & 11 & $10^{-4}$ & 0.19 & 69.3 \\
20 & 0.1 & 0.01 & 12 & $10^{-4}$ & 0.19 & 67.4 \\
21 & 0.1 & 0.01 & 13 & $10^{-4}$ & 0.16 & 57.9 \\
22 & 0.1 & 0.01 & 15 & $10^{-4}$ & 0.03 & 11.7 \\
23 & 0.1 & 0.01 & 16 & $10^{-4}$ & 0 & 0 \\
\hline
24 & 0.1 & 0.01 & 10 & $10^{-2}$ & 0.23 & 82 \\
25 & 0.1 & 0.01 & 10 & $10^{-3}$ & 0.23 & 82 \\
26 & 0.1 & 0.01 & 10 & $10^{-5}$ & 0.04 & 13 \\
27 & 0.1 & 0.01 & 10 & $10^{-6}$ & 0.004 & 1.5 \\
\hline\hline
\end{tabular}
\label{table:allruns} 
\tablefoot{\tablefoottext{a}{In m s$^{-1}$,} \tablefoottext{b}{in Earth masses}.}       
\end{table}

\subsection{Dependence on planetesimal formation efficiency}\label{sub:zeta}

Equation (\ref{sigmaplts}) connects planetesimal formation to evolution of the surface density of pebbles. The parameter $\zeta$ determines what fraction of pebbles is turned to planetesimals within one orbital period. 

Recent local hydrodynamical simulations of streaming instability presented by \citet{2016ApJ...822...55S} report that planetesimal formation saturates on a timescale of a few tens orbital periods at a level of $\sim$50\% of pebbles being turned to planetesimals. This would correspond to $\zeta \approx 10^{-2}$. However, these models used pebbles corresponding to ${\rm St}=0.3$, order of magnitude larger than considered in this paper. For smaller pebbles, the timescale of streaming instability increases, which forces us to also consider lower $\zeta$ values.

We tested a wide range of values for this planetesimal formation efficiency parameter. Naturally, higher $\zeta$ values lead to more planetesimals being formed. The width of the planetesimal annulus slightly changes, reaching a minimum for the highest value of $\zeta=10^{-2}$. Setting $\zeta$ to values even higher than $10^{-2}$ leads to numerical instabilities since too great a mass of dust is removed in one timestep. The mass of planetesimals that form naturally increases with increasing $\zeta$ and saturates at 82 Earth masses for $\zeta=10^{-3}$. For lower $\zeta$ values, the radial drift competes with planetesimal formation in taking the pebbles away from the pile-up region.

\begin{figure*}
   \centering
   \includegraphics[width=\hsize]{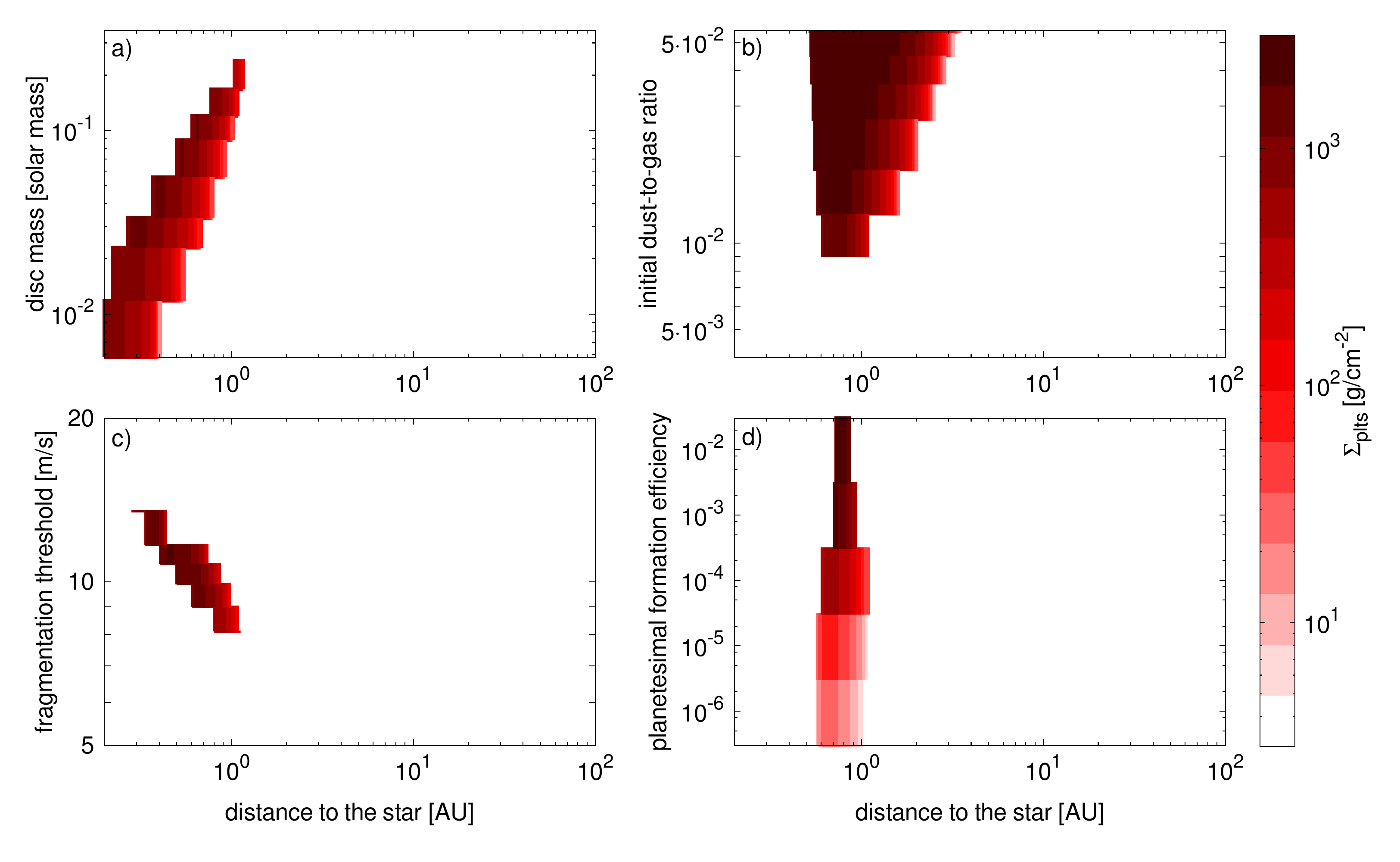}
   \caption{Surface density of planetesimals formed in models with different initial a)~disc mass, b)~dust-to-gas ratio, c)~fragmentation threshold, d)~planetesimal formation efficiency. All the panels share a common colour map for convenience.}
              \label{fig:4panels}%
\end{figure*}

\section{Discussion}\label{sub:discussion}

Our results show that global redistribution of solids driven by dust growth and radial drift leads to a local pile-up of pebbles in the inner, fragmentation dominated part of the protoplanetary disc. This pile-up generates conditions necessary for planetesimal formation. Because the planetesimal formation may only be triggered under such specific conditions, planetesimals do not form at every orbital distance.
This is qualitatively consistent with the latest observations of protoplanetary discs, which suggest that axisymmetric pile-ups of pebbles occur at various distances from the central star \citep{2015ApJ...808L...3A,2016ApJ...820L..40A}. In the models presented in this paper, we do not include processes, such as zonal flows that may potentially create pressure bumps in the outer parts of the disc \citep{2015A&A...574A..68F}. Without these bumps, the pebbles pile-up in the inner part of the disc, where their subsequent evolution is driven by fragmentation rather than radial drift.   

Planetesimal formation is not very efficient in our models, typically incorporating about 20\% of the total solids. However, as this mass is packed into a relatively narrow annulus, the surface density of planetesimals is fairly high, exceeding that predicted by MMSN. The total mass of planetesimals ranges from 1 Earth mass to above 1000 Earth masses in different models. However, we do not know how much mass could actually end up in the final planets - this depends on the efficiency of the late stages of planetary accretion. 

Implications of this local, close-in planetesimal formation for the architectures of planetary systems will be the subject of our future work. Our scenario has a promising potential for explaining some of the solar system features. 
\citet{2009Icar..203..644R} show that it is impossible to reproduce the architecture of the inner solar system starting from an initially uniform distribution of planetesimals. Mars analogues produced in these models were always significantly more massive than the actual Mars. The low mass ratio between Mars and Earth suggests that a significant depletion of solids outside of 1~AU existed before the final assembly of the terrestrial planets \citep{2009ApJ...703.1131H,2014ApJ...782...31I,2015MNRAS.453.3619I}. One possible explanation for this depletion is that it was sculpted by the migration of Jupiter and Saturn, which is known as the Grand Tack scenario \citep{2011Natur.475..206W}. Our results produce initial conditions necessary to reproduce the masses of terrestrial planets in a much more straightforward way.

Another solution to the low mass of the Mars problem is offered by the so-called pebble accretion scenario. Pebbles corresponding to Stokes number of $10^{-2}$-1 may be accreted very efficiently onto planetary embryos \citep{2010A&A...520A..43O,2012A&A...544A..32L}. However, the growth efficiency depends on the embryo size and its location in the disc such that there is a steep cutoff between embryos that can and cannot grow \citep{2015PNAS..11214180L}. The rate of the pebble accretion depends also on the flux and size of accreting pebbles. Large pebbles would lead to a very efficient growth of embryos and formation of too many planets in the terrestrial planet region \citep{2014AJ....148..109K}. Recently, \citet{2015SciA....115109J} present more realistic models with pebbles corresponding to Stokes numbers more similar to those suggested by our models. In this case, the pebble accretion contributes to the growth of terrestrial planets at the same rate as planetesimal accretion.

Our models indicate the formation of a significant pile-up of pebbles close to the central star that then allows for planetesimal formation via the streaming instability.
These pile-ups in the inner disc were previously reported in the dust evolution models of \citet{2002ApJ...580..494Y}, \citet{2004ApJ...601.1109Y}, and \citet{2012A&A...539A.148B}.
A global transition from shallow to steep surface density profiles of pebbles was also observed recently in the TW Hydrae disc \citep{2016A&A...586A..99H}.
The universality of pebble pile-ups would support the in situ formation scenario for the tightly packed exoplanetary systems detected by the {\it Kepler} mission \citep{2012ApJ...761...92F}.

In our work, we use a relatively simple 1D model of a viscous protoplanetary disc. However, many of the observed discs do not appear smooth or axisymmetric \citep{2012ApJ...760L..26M, 2012A&A...546A..24R,2015A&A...578L...6B}, and large fraction of discs exhibit an extended inner cavity \citep{2011ApJ...732...42A}. The peculiar shapes of observed protoplanetary and transition discs may be explained by planet-disc interactions \citep[e.g.][]{2015MNRAS.453.1768P, 2016A&A...585A..35P}. It is also known that presence of planet-induced gaps in discs facilitates the formation of further planets \citep{2012ApJ...756...70K,2015A&A...580A.105P}. Thus, it is formation of the first planet that is most problematic to explain and this is the problem we have considered, and using a simple disc model is justified by the uncertainty of the disc morphology during the very early stages of evolution. 

The shallow gas disc profile we employ is consistent with observations of circumstellar discs \citep{2007ApJ...659..705A}. \citet{2012A&A...539A.148B} notice that the steeper surface density profile $\Sigma_{\rm d}\propto r^{-3/2}$, consistent with the MMSN, is naturally obtained by the dust component during its evolution in the fragmentation dominated regime, even in a primarily shallow disc, as long as the turbulence level is low. This means that the initially constant dust-to-gas ratio changes and it can be enhanced in the inner regions of a shallow disc. In a steep disc, such as the MMSN with $\Sigma_{\rm g}\propto r^{-3/2}$, this type of pile-up does not occur and the inner part of the disc is gradually depleted with the dust-to-gas ratio remaining independent of radial distance. As shown in Fig.~\ref{fig:6panels}, the disc becomes more and more drift-dominated during its evolution. In the drift-dominated regime the surface density of dust evolves to $\Sigma_{\rm d}\propto r^{-3/4}$, which was also found by \citet{2014A&A...572A.107L}.

As mentioned in Sect.~\ref{sub:gasdisc}, we do not include disc heating by stellar irradiation. The midplane temperature drops to $\sim$10~K at large distances from the star, close to the assumed temperature of interstellar medium. By way of contrast, models including irradiation from the star predict higher temperatures \citep[see, e.g.][]{2015A&A...575A..28B}, but these models assume that the accretion rate in the disc is uniform, which is correct close to the star, but questionable at larger distances. In our model, we do not make any assumptions on the time evolution of the accretion rate, but we rather assumed some value for the viscosity (or more precisely the $\alpha_{\rm t}$ parameter) and a value of the mass loss associated with photoevaporation (see Sect.~\ref{sub:gasdisc}). We leave testing another disc models for future work.

A caveat of our models is that we do not explicitly account for the impact of dust on the global evolution of the gas disc, which may be important when the local dust-to-gas ratio reaches unity, since the gas rotation would be modified in this case \citep{2016A&A...591A..86T}. However, since we are mostly interested in obtaining the conditions necessary for triggering the planetesimal formation, before the dust-to-gas ratio reaches such a high value, this is a justified approach.

In this paper, we focus on compact growth during which the internal density of aggregates remains constant. Porous growth may significantly influence the fragmentation and radial drift barriers \citep{2012ApJ...752..106O, 2013A&A...557L...4K} is well known. Recently, \citet{2016A&A...586A..20K} investigated the global evolution of porous aggregates using another simplified method. They find that direct planetesimal growth is hindered by erosion but the sizes of dust aggregates might be sufficient for triggering the streaming instability. However, they did not find significant pile-ups of material as we do, but this is because they only adopted the MMSN model in which the pile-up does not occur as we explained earlier in this section.

We found that fragmentation threshold has to be above 8~m~s$^{-1}$ to allow us for the formation of sufficiently large pebbles in our fiducial disc. These high fragmentation thresholds can be obtained for porous silicate aggregates \citep{2013MNRAS.435.2371M}, preferably built from small monomers as the fragmentation velocity strongly depends on the monomer size \citep{2013A&A...559A..62W}. Thus, the bouncing that is found in laboratory experiments that consider the silicate grains \citep{2010A&A...513A..56G, 2014ApJ...783..111K}, has to be excluded to allow planetesimal formation. This means that planetesimals can only form if grains are relatively sticky. However our models also suggest that the grains cannot grow too large to allow for the pile-up formation (see Sect.~\ref{sub:vth}). The so-called stickiness of dust aggregates is strongly connected to their constituent material: icy grains are considered to be significantly more sticky than silicates \citep{2014MNRAS.437..690A} and thus the fragmentation threshold velocity should depend on the local temperature. We do not include these effects in this study. However, recent models show that if the threshold velocity at which the growth stalls decreases inward to the snow line, a pile-up of solids arises around this location \citep{2015ApJ...815L..15B,2016ApJ...818..200E}, which might be another mechanism for triggering the planetesimal formation. We leave the investigation of this effect for future studies.

\section{Conclusions}\label{sub:last}

This paper addresses the connection between dust evolution and planetesimal formation, which represents a major gap in state-of-the-art planet formation models. As dust growth is limited by fragmentation and radial drift, a direct growth to planetesimal sizes seems unlikely. However, the same processes drive a global redistribution of solids and may lead to a pile-up of pebbles that triggers planetesimal formation via the streaming instability. 
 
We show that a narrow ring of planetesimals with a steep surface density profile is naturally produced in the inner part of a shallow protoplanetary disc corresponding to the observed ones. 
These planetesimals form from the flux of solids originating from the outer disc that are carried by the radial drift. 
At the same time, the radial drift limits the radial extent of this annulus, as it prevents the dust-to-gas ratio enhancement from spreading further out. On the other hand, the inner edge is caused by lack of sufficiently large pebbles that could trigger the streaming instability. Impact velocities increase towards the inner disc because of higher temperatures and thus the maximum size of grains decreases. The exact properties of planetesimal annulus depend on various parameters, as shown in Fig.~\ref{fig:4panels} and summarised in Table~\ref{table:allruns}.

The narrow annulus of planetesimals around 1~AU enables us to reproduce the unusual masses of terrestrial planets, where Earth and Venus are significantly more massive than Mercury and Mars \citep{2009ApJ...703.1131H}. In some of our models, the planetesimal annulus contains as much as 1 000 Earth masses, which could allow for in situ formation of the close-in massive planets that have been detected around many stars \citep{2013ApJ...766...81F}.

Our results, including the surface density of planetesimals and timing of their formation, may be used as an input to models which investigate the later stages of planet accretion, planet population synthesis, and the internal evolution of asteroids. The size and flux of pebbles that we obtain is important for the pebble accretion models that typically assume unphysically large aggregates and make ad hoc assumptions about their surface density.    

\begin{acknowledgements}
     The authors thank Bertram Bitsch, Kees Dullemond, Barbara Ercolano, Anders Johansen, and Chris Ormel for useful discussions.
     We also thank the referee for their useful report that helped us to significantly improve this paper. 
     This work has been carried out within the frame of the PlanetS National Center of Competence in Research (NCCR) 
     of the Swiss National Science Foundation. The numerical models were performed on the zBox4 CPU cluster 
     at the Institute for Computational Science, University of Zurich.
\end{acknowledgements}

\bibliographystyle{aa} 
\bibliography{paper.bib}

\begin{thebibliography}{75}
\expandafter\ifx\csname natexlab\endcsname\relax\def\natexlab#1{#1}\fi

\bibitem[{{Alibert} {et~al.}(2013){Alibert}, {Carron}, {Fortier}, {Pfyffer},
  {Benz}, {Mordasini}, \& {Swoboda}}]{2013A&A...558A.109A}
{Alibert}, Y., {Carron}, F., {Fortier}, A., {et~al.} 2013, \aap, 558, A109

\bibitem[{{Alibert} {et~al.}(2005){Alibert}, {Mordasini}, {Benz}, \&
  {Winisdoerffer}}]{2005A&A...434..343A}
{Alibert}, Y., {Mordasini}, C., {Benz}, W., \& {Winisdoerffer}, C. 2005, \aap,
  434, 343

\bibitem[{{ALMA Partnership} {et~al.}(2015){ALMA Partnership}, {Brogan},
  {P{\'e}rez}, {Hunter}, {Dent}, {Hales}, {Hills}, {Corder}, {Fomalont},
  {Vlahakis}, {Asaki}, {Barkats}, {Hirota}, {Hodge}, {Impellizzeri}, {Kneissl},
  {Liuzzo}, {Lucas}, {Marcelino}, {Matsushita}, {Nakanishi}, {Phillips},
  {Richards}, {Toledo}, {Aladro}, {Broguiere}, {Cortes}, {Cortes}, {Espada},
  {Galarza}, {Garcia-Appadoo}, {Guzman-Ramirez}, {Humphreys}, {Jung}, {Kameno},
  {Laing}, {Leon}, {Marconi}, {Mignano}, {Nikolic}, {Nyman}, {Radiszcz},
  {Remijan}, {Rod{\'o}n}, {Sawada}, {Takahashi}, {Tilanus}, {Vila Vilaro},
  {Watson}, {Wiklind}, {Akiyama}, {Chapillon}, {de Gregorio-Monsalvo}, {Di
  Francesco}, {Gueth}, {Kawamura}, {Lee}, {Nguyen Luong}, {Mangum}, {Pietu},
  {Sanhueza}, {Saigo}, {Takakuwa}, {Ubach}, {van Kempen}, {Wootten},
  {Castro-Carrizo}, {Francke}, {Gallardo}, {Garcia}, {Gonzalez}, {Hill},
  {Kaminski}, {Kurono}, {Liu}, {Lopez}, {Morales}, {Plarre}, {Schieven},
  {Testi}, {Videla}, {Villard}, {Andreani}, {Hibbard}, \&
  {Tatematsu}}]{2015ApJ...808L...3A}
{ALMA Partnership}, {Brogan}, C.~L., {P{\'e}rez}, L.~M., {et~al.} 2015, \apjl,
  808, L3

\bibitem[{{Andrews} \& {Williams}(2007)}]{2007ApJ...659..705A}
{Andrews}, S.~M. \& {Williams}, J.~P. 2007, \apj, 659, 705

\bibitem[{{Andrews} {et~al.}(2011){Andrews}, {Wilner}, {Espaillat}, {Hughes},
  {Dullemond}, {McClure}, {Qi}, \& {Brown}}]{2011ApJ...732...42A}
{Andrews}, S.~M., {Wilner}, D.~J., {Espaillat}, C., {et~al.} 2011, \apj, 732,
  42

\bibitem[{{Andrews} {et~al.}(2016){Andrews}, {Wilner}, {Zhu}, {Birnstiel},
  {Carpenter}, {P{\'e}rez}, {Bai}, {{\"O}berg}, {Hughes}, {Isella}, \&
  {Ricci}}]{2016ApJ...820L..40A}
{Andrews}, S.~M., {Wilner}, D.~J., {Zhu}, Z., {et~al.} 2016, \apjl, 820, L40

\bibitem[{{Armitage}(2013)}]{2013apf..book.....A}
{Armitage}, P.~J. 2013, {Astrophysics of Planet Formation}

\bibitem[{{Aumatell} \& {Wurm}(2014)}]{2014MNRAS.437..690A}
{Aumatell}, G. \& {Wurm}, G. 2014, \mnras, 437, 690

\bibitem[{{Ayliffe} {et~al.}(2012){Ayliffe}, {Laibe}, {Price}, \&
  {Bate}}]{2012MNRAS.423.1450A}
{Ayliffe}, B.~A., {Laibe}, G., {Price}, D.~J., \& {Bate}, M.~R. 2012, \mnras,
  423, 1450

\bibitem[{{Bai} \& {Stone}(2010{\natexlab{a}})}]{2010ApJ...722.1437B}
{Bai}, X.-N. \& {Stone}, J.~M. 2010{\natexlab{a}}, \apj, 722, 1437

\bibitem[{{Bai} \& {Stone}(2010{\natexlab{b}})}]{2010ApJ...722L.220B}
{Bai}, X.-N. \& {Stone}, J.~M. 2010{\natexlab{b}}, \apjl, 722, L220

\bibitem[{{Banzatti} {et~al.}(2015){Banzatti}, {Pinilla}, {Ricci},
  {Pontoppidan}, {Birnstiel}, \& {Ciesla}}]{2015ApJ...815L..15B}
{Banzatti}, A., {Pinilla}, P., {Ricci}, L., {et~al.} 2015, \apjl, 815, L15

\bibitem[{{Bell} \& {Lin}(1994)}]{1994ApJ...427..987B}
{Bell}, K.~R. \& {Lin}, D.~N.~C. 1994, \apj, 427, 987

\bibitem[{{Benisty} {et~al.}(2015){Benisty}, {Juhasz}, {Boccaletti},
  {Avenhaus}, {Milli}, {Thalmann}, {Dominik}, {Pinilla}, {Buenzli}, {Pohl},
  {Beuzit}, {Birnstiel}, {de Boer}, {Bonnefoy}, {Chauvin}, {Christiaens},
  {Garufi}, {Grady}, {Henning}, {Huelamo}, {Isella}, {Langlois}, {M{\'e}nard},
  {Mouillet}, {Olofsson}, {Pantin}, {Pinte}, \& {Pueyo}}]{2015A&A...578L...6B}
{Benisty}, M., {Juhasz}, A., {Boccaletti}, A., {et~al.} 2015, \aap, 578, L6

\bibitem[{{Birnstiel} {et~al.}(2009){Birnstiel}, {Dullemond}, \&
  {Brauer}}]{2009A&A...503L...5B}
{Birnstiel}, T., {Dullemond}, C.~P., \& {Brauer}, F. 2009, \aap, 503, L5

\bibitem[{{Birnstiel} {et~al.}(2012){Birnstiel}, {Klahr}, \&
  {Ercolano}}]{2012A&A...539A.148B}
{Birnstiel}, T., {Klahr}, H., \& {Ercolano}, B. 2012, \aap, 539, A148

\bibitem[{{Birnstiel} {et~al.}(2011){Birnstiel}, {Ormel}, \&
  {Dullemond}}]{2011A&A...525A..11B}
{Birnstiel}, T., {Ormel}, C.~W., \& {Dullemond}, C.~P. 2011, \aap, 525, A11

\bibitem[{{Bitsch} {et~al.}(2015){Bitsch}, {Johansen}, {Lambrechts}, \&
  {Morbidelli}}]{2015A&A...575A..28B}
{Bitsch}, B., {Johansen}, A., {Lambrechts}, M., \& {Morbidelli}, A. 2015, \aap,
  575, A28

\bibitem[{{Brauer} {et~al.}(2008){Brauer}, {Dullemond}, \&
  {Henning}}]{2008A&A...480..859B}
{Brauer}, F., {Dullemond}, C.~P., \& {Henning}, T. 2008, \aap, 480, 859

\bibitem[{{Carrera} {et~al.}(2015){Carrera}, {Johansen}, \&
  {Davies}}]{2015A&A...579A..43C}
{Carrera}, D., {Johansen}, A., \& {Davies}, M.~B. 2015, \aap, 579, A43

\bibitem[{{Cassan} {et~al.}(2012){Cassan}, {Kubas}, {Beaulieu}, {Dominik},
  {Horne}, {Greenhill}, {Wambsganss}, {Menzies}, {Williams}, {J{\o}rgensen},
  {Udalski}, {Bennett}, {Albrow}, {Batista}, {Brillant}, {Caldwell}, {Cole},
  {Coutures}, {Cook}, {Dieters}, {Prester}, {Donatowicz}, {Fouqu{\'e}}, {Hill},
  {Kains}, {Kane}, {Marquette}, {Martin}, {Pollard}, {Sahu}, {Vinter},
  {Warren}, {Watson}, {Zub}, {Sumi}, {Szyma{\'n}ski}, {Kubiak}, {Poleski},
  {Soszynski}, {Ulaczyk}, {Pietrzy{\'n}ski}, \&
  {Wyrzykowski}}]{2012Natur.481..167C}
{Cassan}, A., {Kubas}, D., {Beaulieu}, J.-P., {et~al.} 2012, \nat, 481, 167

\bibitem[{{Dittrich} {et~al.}(2013){Dittrich}, {Klahr}, \&
  {Johansen}}]{2013ApJ...763..117D}
{Dittrich}, K., {Klahr}, H., \& {Johansen}, A. 2013, \apj, 763, 117

\bibitem[{{Dr{\c a}{\.z}kowska} \& {Dullemond}(2014)}]{2014A&A...572A..78D}
{Dr{\c a}{\.z}kowska}, J. \& {Dullemond}, C.~P. 2014, \aap, 572, A78

\bibitem[{{Estrada} {et~al.}(2016){Estrada}, {Cuzzi}, \&
  {Morgan}}]{2016ApJ...818..200E}
{Estrada}, P.~R., {Cuzzi}, J.~N., \& {Morgan}, D.~A. 2016, \apj, 818, 200

\bibitem[{{Fang} \& {Margot}(2012)}]{2012ApJ...761...92F}
{Fang}, J. \& {Margot}, J.-L. 2012, \apj, 761, 92

\bibitem[{{Flock} {et~al.}(2015){Flock}, {Ruge}, {Dzyurkevich}, {Henning},
  {Klahr}, \& {Wolf}}]{2015A&A...574A..68F}
{Flock}, M., {Ruge}, J.~P., {Dzyurkevich}, N., {et~al.} 2015, \aap, 574, A68

\bibitem[{{Fressin} {et~al.}(2013){Fressin}, {Torres}, {Charbonneau}, {Bryson},
  {Christiansen}, {Dressing}, {Jenkins}, {Walkowicz}, \&
  {Batalha}}]{2013ApJ...766...81F}
{Fressin}, F., {Torres}, G., {Charbonneau}, D., {et~al.} 2013, \apj, 766, 81

\bibitem[{{G{\"u}ttler} {et~al.}(2010){G{\"u}ttler}, {Blum}, {Zsom}, {Ormel},
  \& {Dullemond}}]{2010A&A...513A..56G}
{G{\"u}ttler}, C., {Blum}, J., {Zsom}, A., {Ormel}, C.~W., \& {Dullemond},
  C.~P. 2010, \aap, 513, A56

\bibitem[{{Hansen}(2009)}]{2009ApJ...703.1131H}
{Hansen}, B.~M.~S. 2009, \apj, 703, 1131

\bibitem[{{Hogerheijde} {et~al.}(2016){Hogerheijde}, {Bekkers}, {Pinilla},
  {Salinas}, {Kama}, {Andrews}, {Qi}, \& {Wilner}}]{2016A&A...586A..99H}
{Hogerheijde}, M.~R., {Bekkers}, D., {Pinilla}, P., {et~al.} 2016, \aap, 586,
  A99

\bibitem[{{Izidoro} {et~al.}(2014){Izidoro}, {Haghighipour}, {Winter}, \&
  {Tsuchida}}]{2014ApJ...782...31I}
{Izidoro}, A., {Haghighipour}, N., {Winter}, O.~C., \& {Tsuchida}, M. 2014,
  \apj, 782, 31

\bibitem[{{Izidoro} {et~al.}(2015){Izidoro}, {Raymond}, {Morbidelli}, \&
  {Winter}}]{2015MNRAS.453.3619I}
{Izidoro}, A., {Raymond}, S.~N., {Morbidelli}, A., \& {Winter}, O.~C. 2015,
  \mnras, 453, 3619

\bibitem[{{Johansen} {et~al.}(2011){Johansen}, {Klahr}, \&
  {Henning}}]{2011A&A...529A..62J}
{Johansen}, A., {Klahr}, H., \& {Henning}, T. 2011, \aap, 529, A62

\bibitem[{{Johansen} {et~al.}(2015){Johansen}, {Mac Low}, {Lacerda}, \&
  {Bizzarro}}]{2015SciA....115109J}
{Johansen}, A., {Mac Low}, M.-M., {Lacerda}, P., \& {Bizzarro}, M. 2015,
  Science Advances, 1, 1500109

\bibitem[{{Johansen} {et~al.}(2007){Johansen}, {Oishi}, {Mac Low}, {Klahr},
  {Henning}, \& {Youdin}}]{2007Natur.448.1022J}
{Johansen}, A., {Oishi}, J.~S., {Mac Low}, M.-M., {et~al.} 2007, \nat, 448,
  1022

\bibitem[{{Johansen} {et~al.}(2009){Johansen}, {Youdin}, \& {Mac
  Low}}]{2009ApJ...704L..75J}
{Johansen}, A., {Youdin}, A., \& {Mac Low}, M.-M. 2009, \apjl, 704, L75

\bibitem[{{Kataoka} {et~al.}(2013){Kataoka}, {Tanaka}, {Okuzumi}, \&
  {Wada}}]{2013A&A...557L...4K}
{Kataoka}, A., {Tanaka}, H., {Okuzumi}, S., \& {Wada}, K. 2013, \aap, 557, L4

\bibitem[{{Kelling} {et~al.}(2014){Kelling}, {Wurm}, \&
  {K{\"o}ster}}]{2014ApJ...783..111K}
{Kelling}, T., {Wurm}, G., \& {K{\"o}ster}, M. 2014, \apj, 783, 111

\bibitem[{{Kobayashi} {et~al.}(2012){Kobayashi}, {Ormel}, \&
  {Ida}}]{2012ApJ...756...70K}
{Kobayashi}, H., {Ormel}, C.~W., \& {Ida}, S. 2012, \apj, 756, 70

\bibitem[{{Kretke} \& {Levison}(2014)}]{2014AJ....148..109K}
{Kretke}, K.~A. \& {Levison}, H.~F. 2014, \aj, 148, 109

\bibitem[{{Krijt} {et~al.}(2016){Krijt}, {Ormel}, {Dominik}, \&
  {Tielens}}]{2016A&A...586A..20K}
{Krijt}, S., {Ormel}, C.~W., {Dominik}, C., \& {Tielens}, A.~G.~G.~M. 2016,
  \aap, 586, A20

\bibitem[{{Laibe} {et~al.}(2012){Laibe}, {Gonzalez}, \&
  {Maddison}}]{2012A&A...537A..61L}
{Laibe}, G., {Gonzalez}, J.-F., \& {Maddison}, S.~T. 2012, \aap, 537, A61

\bibitem[{{Lambrechts} \& {Johansen}(2012)}]{2012A&A...544A..32L}
{Lambrechts}, M. \& {Johansen}, A. 2012, \aap, 544, A32

\bibitem[{{Lambrechts} \& {Johansen}(2014)}]{2014A&A...572A.107L}
{Lambrechts}, M. \& {Johansen}, A. 2014, \aap, 572, A107

\bibitem[{{Levison} {et~al.}(2015){Levison}, {Kretke}, {Walsh}, \&
  {Bottke}}]{2015PNAS..11214180L}
{Levison}, H.~F., {Kretke}, K.~A., {Walsh}, K.~J., \& {Bottke}, W.~F. 2015,
  Proceedings of the National Academy of Science, 112, 14180

\bibitem[{{Lyra} {et~al.}(2009){Lyra}, {Johansen}, {Zsom}, {Klahr}, \&
  {Piskunov}}]{2009A&A...497..869L}
{Lyra}, W., {Johansen}, A., {Zsom}, A., {Klahr}, H., \& {Piskunov}, N. 2009,
  \aap, 497, 869

\bibitem[{{Mayama} {et~al.}(2012){Mayama}, {Hashimoto}, {Muto}, {Tsukagoshi},
  {Kusakabe}, {Kuzuhara}, {Takahashi}, {Kudo}, {Dong}, {Fukagawa}, {Takami},
  {Momose}, {Wisniewski}, {Follette}, {Abe}, {Akiyama}, {Brandner}, {Brandt},
  {Carson}, {Egner}, {Feldt}, {Goto}, {Grady}, {Guyon}, {Hayano}, {Hayashi},
  {Hayashi}, {Henning}, {Hodapp}, {Ishii}, {Iye}, {Janson}, {Kandori}, {Kwon},
  {Knapp}, {Matsuo}, {McElwain}, {Miyama}, {Morino}, {Moro-Martin},
  {Nishimura}, {Pyo}, {Serabyn}, {Suto}, {Suzuki}, {Takato}, {Terada},
  {Thalmann}, {Tomono}, {Turner}, {Watanabe}, {Yamada}, {Takami}, {Usuda}, \&
  {Tamura}}]{2012ApJ...760L..26M}
{Mayama}, S., {Hashimoto}, J., {Muto}, T., {et~al.} 2012, \apjl, 760, L26

\bibitem[{{Meru} {et~al.}(2013){Meru}, {Geretshauser}, {Sch{\"a}fer}, {Speith},
  \& {Kley}}]{2013MNRAS.435.2371M}
{Meru}, F., {Geretshauser}, R.~J., {Sch{\"a}fer}, C., {Speith}, R., \& {Kley},
  W. 2013, \mnras, 435, 2371

\bibitem[{{Nakagawa} {et~al.}(1986){Nakagawa}, {Sekiya}, \&
  {Hayashi}}]{1986Icar...67..375N}
{Nakagawa}, Y., {Sekiya}, M., \& {Hayashi}, C. 1986, \icarus, 67, 375

\bibitem[{{Okuzumi} {et~al.}(2012){Okuzumi}, {Tanaka}, {Kobayashi}, \&
  {Wada}}]{2012ApJ...752..106O}
{Okuzumi}, S., {Tanaka}, H., {Kobayashi}, H., \& {Wada}, K. 2012, \apj, 752,
  106

\bibitem[{{Ormel} \& {Cuzzi}(2007)}]{2007A&A...466..413O}
{Ormel}, C.~W. \& {Cuzzi}, J.~N. 2007, \aap, 466, 413

\bibitem[{{Ormel} \& {Klahr}(2010)}]{2010A&A...520A..43O}
{Ormel}, C.~W. \& {Klahr}, H.~H. 2010, \aap, 520, A43

\bibitem[{{Papaloizou} \& {Terquem}(1999)}]{1999ApJ...521..823P}
{Papaloizou}, J.~C.~B. \& {Terquem}, C. 1999, \apj, 521, 823

\bibitem[{{Pinilla} {et~al.}(2015){Pinilla}, {Birnstiel}, \&
  {Walsh}}]{2015A&A...580A.105P}
{Pinilla}, P., {Birnstiel}, T., \& {Walsh}, C. 2015, \aap, 580, A105

\bibitem[{{Pinilla} {et~al.}(2016){Pinilla}, {Klarmann}, {Birnstiel},
  {Benisty}, {Dominik}, \& {Dullemond}}]{2016A&A...585A..35P}
{Pinilla}, P., {Klarmann}, L., {Birnstiel}, T., {et~al.} 2016, \aap, 585, A35

\bibitem[{{Pinte} \& {Laibe}(2014)}]{2014A&A...565A.129P}
{Pinte}, C. \& {Laibe}, G. 2014, \aap, 565, A129

\bibitem[{{Pohl} {et~al.}(2015){Pohl}, {Pinilla}, {Benisty}, {Ataiee},
  {Juh{\'a}sz}, {Dullemond}, {Van Boekel}, \& {Henning}}]{2015MNRAS.453.1768P}
{Pohl}, A., {Pinilla}, P., {Benisty}, M., {et~al.} 2015, \mnras, 453, 1768

\bibitem[{{Raettig} {et~al.}(2015){Raettig}, {Klahr}, \&
  {Lyra}}]{2015ApJ...804...35R}
{Raettig}, N., {Klahr}, H., \& {Lyra}, W. 2015, \apj, 804, 35

\bibitem[{{Rameau} {et~al.}(2012){Rameau}, {Chauvin}, {Lagrange},
  {Th{\'e}bault}, {Milli}, {Girard}, \& {Bonnefoy}}]{2012A&A...546A..24R}
{Rameau}, J., {Chauvin}, G., {Lagrange}, A.-M., {et~al.} 2012, \aap, 546, A24

\bibitem[{{Raymond} {et~al.}(2009){Raymond}, {O'Brien}, {Morbidelli}, \&
  {Kaib}}]{2009Icar..203..644R}
{Raymond}, S.~N., {O'Brien}, D.~P., {Morbidelli}, A., \& {Kaib}, N.~A. 2009,
  \icarus, 203, 644

\bibitem[{{Shakura} \& {Sunyaev}(1973)}]{1973A&A....24..337S}
{Shakura}, N.~I. \& {Sunyaev}, R.~A. 1973, \aap, 24, 337

\bibitem[{{Simon} {et~al.}(2016){Simon}, {Armitage}, {Li}, \&
  {Youdin}}]{2016ApJ...822...55S}
{Simon}, J.~B., {Armitage}, P.~J., {Li}, R., \& {Youdin}, A.~N. 2016, \apj,
  822, 55

\bibitem[{{Surville} {et~al.}(2016){Surville}, {Mayer}, \&
  {Lin}}]{2016arXiv160105945S}
{Surville}, C., {Mayer}, L., \& {Lin}, D.~N.~C. 2016, ArXiv e-prints
  [\eprint[arXiv]{1601.05945}]

\bibitem[{{Taki} {et~al.}(2016){Taki}, {Fujimoto}, \&
  {Ida}}]{2016A&A...591A..86T}
{Taki}, T., {Fujimoto}, M., \& {Ida}, S. 2016, \aap, 591, A86

\bibitem[{{Tanaka} {et~al.}(2005){Tanaka}, {Himeno}, \&
  {Ida}}]{2005ApJ...625..414T}
{Tanaka}, H., {Himeno}, Y., \& {Ida}, S. 2005, \apj, 625, 414

\bibitem[{{Testi} {et~al.}(2014){Testi}, {Birnstiel}, {Ricci}, {Andrews},
  {Blum}, {Carpenter}, {Dominik}, {Isella}, {Natta}, {Williams}, \&
  {Wilner}}]{2014prpl.conf..339T}
{Testi}, L., {Birnstiel}, T., {Ricci}, L., {et~al.} 2014, Protostars and
  Planets VI, 339

\bibitem[{{Udry} \& {Santos}(2007)}]{2007ARA&A..45..397U}
{Udry}, S. \& {Santos}, N.~C. 2007, \araa, 45, 397

\bibitem[{{Veras} \& {Armitage}(2004)}]{2004MNRAS.347..613V}
{Veras}, D. \& {Armitage}, P.~J. 2004, \mnras, 347, 613

\bibitem[{{Wada} {et~al.}(2013){Wada}, {Tanaka}, {Okuzumi}, {Kobayashi},
  {Suyama}, {Kimura}, \& {Yamamoto}}]{2013A&A...559A..62W}
{Wada}, K., {Tanaka}, H., {Okuzumi}, S., {et~al.} 2013, \aap, 559, A62

\bibitem[{{Walsh} {et~al.}(2011){Walsh}, {Morbidelli}, {Raymond}, {O'Brien}, \&
  {Mandell}}]{2011Natur.475..206W}
{Walsh}, K.~J., {Morbidelli}, A., {Raymond}, S.~N., {O'Brien}, D.~P., \&
  {Mandell}, A.~M. 2011, \nat, 475, 206

\bibitem[{{Weidenschilling}(1977)}]{1977MNRAS.180...57W}
{Weidenschilling}, S.~J. 1977, \mnras, 180, 57

\bibitem[{{Weidenschilling}(1997)}]{1997Icar..127..290W}
{Weidenschilling}, S.~J. 1997, \icarus, 127, 290

\bibitem[{{Youdin} \& {Chiang}(2004)}]{2004ApJ...601.1109Y}
{Youdin}, A.~N. \& {Chiang}, E.~I. 2004, \apj, 601, 1109

\bibitem[{{Youdin} \& {Shu}(2002)}]{2002ApJ...580..494Y}
{Youdin}, A.~N. \& {Shu}, F.~H. 2002, \apj, 580, 494

\bibitem[{{Zsom} {et~al.}(2010){Zsom}, {Ormel}, {G{\"u}ttler}, {Blum}, \&
  {Dullemond}}]{2010A&A...513A..57Z}
{Zsom}, A., {Ormel}, C.~W., {G{\"u}ttler}, C., {Blum}, J., \& {Dullemond},
  C.~P. 2010, \aap, 513, A57

\end{thebibliography}

\begin{appendix}
\section{Radial drift dependence on dust-to-gas ratio and size distribution}\label{sub:mNSH}

Most of the dust evolution models use the radial drift speed derived by \citet{1977MNRAS.180...57W}
\begin{equation}\label{vr1977}
v_{\rm{r},1977} = \eta v_{\rm{K}} \frac{2\rm{St}}{1 + \rm{St}^{2}},
\end{equation}
where $\eta v_{\rm{K}}$ is the maximum drift speed that results from the gas pressure gradient and $\rm{St}$ is the Stokes number of a particle.
 
Already \citet{1986Icar...67..375N} noticed that the radial drift velocity should be reduced when the solids density is high. They derived the dependence of the drift speed on the dust-to-gas ratio in the case of monodisperse dust population
\begin{equation}\label{vr1986}
v_{\rm{r},1986} = \eta v_{\rm{K}}\frac{2\rm{St}}{\left(1 + \epsilon \right)^2 + \rm{St}^{2}},
\end{equation}
where $\epsilon$ is the local dust-to-gas ratio (of a single sized dust species).

However, in a realistic case, when we have dust particles of different sizes, the dust species interact and exchange the angular momentum leading to drift velocity being dependent on the particular size distribution. In this paper, we implement this effect, which turns out to be of great importance to the dust retention and planetesimal formation (see Fig.~\ref{fig:massevo}). We use the formulation derived by \citet{2005ApJ...625..414T} (their Eqs. 14-18), which can also be found in \citet{2012ApJ...752..106O} (their Eqs. 48-53). The drift velocity of particles with Stokes number $\rm{St}$ is given by
\begin{equation}\label{vmnsh}
v_{\rm{r}}({\rm St}) = \frac{1}{1+\rm{St}^2} v_{\rm{g,r}} + \frac{2\rm{St}}{1+\rm{St}^2} v^{\prime}_{\rm{g,\phi}},
\end{equation}
where the radial velocity of gas is
\begin{equation}
v_{\rm{g,r}} = \frac{2Y}{(1+X)^2+Y^2} \eta v_{\rm{K}},
\end{equation}
and the difference between the azimuthal velocity of gas and the Keplerian velocity is
\begin{equation}
v^{\prime}_{\rm{g,\phi}} = - \frac{1+X}{(1+X)^2+Y^2} \eta v_{\rm{K}}.
\end{equation}
The $X$ and $Y$ are dimensionless quantities dependent on the mass distribution
\begin{equation}
X = \int\epsilon(m)\frac{1}{1+{\rm{St}}(m)^2}dm,
\end{equation}
\begin{equation}
Y = \int\epsilon(m)\frac{{\rm{St}}(m)}{1+{\rm{St}}(m)^2}dm,
\end{equation}
where $\epsilon(m) = \rho_{\rm d}(m) \slash \rho_{\rm g}$.

\begin{figure}
   \centering
   \includegraphics[width=0.95\hsize]{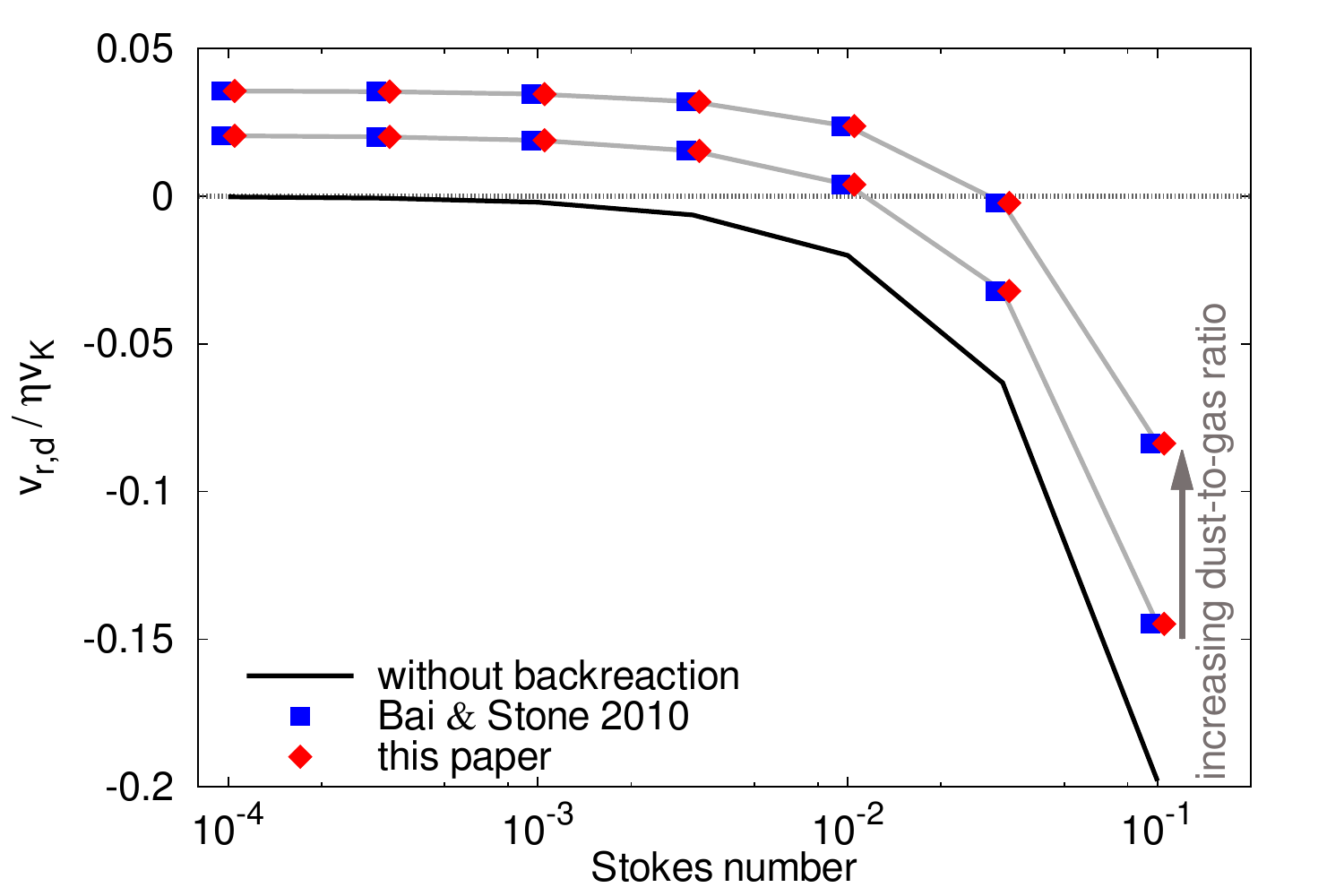}
      \caption{Midplane radial drift speed dependence on the Stokes number for a simple mass distribution and vertically integrated dust-to-gas ratios of 0.01, 0.03, and 0.1. The inward drift is significantly reduced for the higher dust-to-gas ratio. Outward drift is possible for the smallest grains.}
      \label{fig:mNSH}
\end{figure}

We note that the radial drift velocity described by Eq.~(\ref{vmnsh}) naturally converges to the nominal radial drift solution of \citet{1977MNRAS.180...57W} (Eq.~\ref{vr1977}) for low dust-to-gas mass ratios ($\lesssim$ 0.01, the black line in Fig.~\ref{fig:mNSH}) and of \citet{1986Icar...67..375N} (Eq.~\ref{vr1986}) for single sized grains.

The same process was also described by \citet{2010ApJ...722.1437B} (see their Appendix A), who derived a solution based on the matrix formalism (their Eqs. A1-A5 ). Figure \ref{fig:mNSH} shows that these two approaches are equivalent and demonstrate how this effect works for vertically integrated dust-to-gas ratios of 0.01, 0.03, and 0.1 (1, 3, and 10 times the solar value). For this plot, we used a power-law size (and Stokes number) distribution that is an effect of the coagulation-fragmentation equilibrium \citep{2011A&A...525A..11B}
\begin{equation}\label{sizedistr}
n({\rm St})\cdot m\ d{\rm St} \propto {\rm St}^{-1/2}\ d{\rm St}
\end{equation}
for seven equally spaced logarithmic bins between $\rm{St}=10^{-4}$ and $\rm{St}=10^{-1}$. We assumed that there is an equilibrium between settling and turbulent mixing with $\alpha_{\rm t}=10^{-3}$, resulting in the midplane dust-to-gas ratios described by Eq.~(\ref{dtg}). The inward drift of the largest particles is suppressed for the increasing dust-to-gas ratio and even the outward drift ($v_{\rm{r}}>0$) is possible for the smallest grains. The average drift velocity depends both on the total dust-to-gas ratio and mass distribution. 

We assume the size distribution described with Eq.~(\ref{sizedistr}) in the fragmentation regime case in our models. The mass weighted average that is used in Eq.~(\ref{advdiff}) is calculated as
\begin{equation}\label{mwvel}
\bar{v} = \frac{\sum_{\rm St} \epsilon({\rm St}) v_{\rm r}({\rm St})}{\sum_{\rm St} \epsilon({\rm St})}.
\end{equation}
We find that the number of mass bins plays a role in the radial drift velocity calculation. We use 200 bins, since we found the convergence of $\bar{v}$ for this value. 

\section{List of symbols used in the paper}\label{sub:symbols}

In this Appendix, we compile Table~\ref{table:symbols} which describes symbols used throughout this paper and provides their typical value when applicable.

\begin{table}
\caption{List of symbols used in this paper}
\centering                         
\begin{tabular}{l l l}   
\hline\hline                
Symbol & Explanation & Typical value\tablefootmark{a} \\   
\hline
  $r$ & radial distance to the star & \\
  $a$ & particle's radius &    \\
  $a_0$ & monomer size & 1~$\mu$m \\
  $a_{\rm frag}$ & fragmentation limited size &  \\
  $a_{\rm drift}$ & drift limited size &  \\
  $\tau_{\rm growth}$ & growth timescale &  \\
  $\tau_{\rm drift}$ & radial drift timescale &  \\
  $\rm{St}$  & particle's Stokes number &    \\
  $t_{\rm{s}}$ & particle's stopping time & \\
  $\rho_{\bullet}$ & particle's internal density & 1~g~cm$^{-3}$ \\
  $v_{\rm{K}}$ & Keplerian velocity & \\
  $\Omega_{\rm{K}}$ & Keplerian frequency & \\
  $T_{\rm{K}}$ & orbital period & \\
  $\Sigma_{\rm g}$ & surface density of gas & \\
  $\Sigma_{\rm d}$ & surface density of dust & \\
  $\rho_{\rm g}$ & gas density & \\
  $\rho_{\rm d}$ & dust density & \\
  $\epsilon$ & local dust-to-gas ratio & \\
  $\Delta v$ & impact velocity & \\
  $v_{\rm{th}}$ & fragmentation threshold velocity & 10~m~s$^{-1}$ \\
  $v_{\rm{r,d}}$ & radial drift velocity of dust & \\
  $\bar{v}$ & mass weighted average drift speed & \\
  $\eta v_{\rm K}$ & headwind speed & \\
  $H_{\rm{g}}$ & scale height of gas & \\
  $H_{\rm{d}}$ & scale height of dust & \\
  $\alpha_{\rm{t}}$ & turbulence strength & $10^{-3}$\\
  $c_{\rm{s}}$ & sound speed in gas & \\
  $P_{\rm{g}}$ & gas pressure & \\
  $D_{\rm{g}}$ & gas diffusivity & \\
  $Z$ & vertically integrated dust-to-gas ratio & $10^{-2}$ \\
  $M_{\star}$ & mass of the central star & 1~M$_{\odot}$ \\
  $M_{\rm{disc}}$ & protoplanetary disc mass & 0.1~M$_{\odot}$ \\
  $M_{\rm{tot}}$ & total initial mass of solids & \\
  $M_{\rm{plts}}$ & final mass of planetesimals & \\
\hline\hline
\end{tabular}
\label{table:symbols}          
\tablefoot{\tablefoottext{a}{If applicable}.}
\end{table}
\end{appendix}

\end{document}